\documentclass[aps,prl,twocolumn,float]{revtex4-1}
\usepackage[normalem]{ulem}
\usepackage{amsmath}
\usepackage{amssymb}
\usepackage{color}
\usepackage{graphicx}
\usepackage{natbib}
\usepackage{CJK}
\usepackage{lipsum}  
\usepackage{tikz}      
\usepackage[left]{lineno}

\usepackage{xcolor,hyperref}
\hypersetup{
   colorlinks,
   linkcolor={blue!50!black},
   citecolor={blue!50!black},
   urlcolor={blue!80!black}
}
\DeclareMathAlphabet{\mathitbf}{OML}{cmm}{b}{it}

\usepackage{array, booktabs, multirow, xltabular}
\newcommand{\theadl}[1]{\multicolumn{1}{l}{\small\itshape\begin{tabular}{l} #1\end{tabular}}}
\newcommand{\theadc}[1]{\multicolumn{1}{c}{\small\itshape\begin{tabular}{c} #1\end{tabular}}}

\definecolor{pink}{rgb}{0.7,0.05,0.5} 
\definecolor{red}{rgb}{1,0,0} 
\definecolor{blue}{rgb}{0,0,0.7} 
\definecolor{green}{rgb}{0,1,0}
\definecolor{yellow}{rgb}{1,1,0}
\definecolor{orange}{rgb}{1,0.5,0}
\definecolor{white}{rgb}{1,1,1}

\begin{document}

\title{Bridging the Gap Between Avalanche Relaxation and Yielding Rheology }

\author{Leonardo Relmucao-Leiva${}^{1}$, Carlos Villarroel${}^{1}$ and  Gustavo D\"uring${}^{1}$  }
\affiliation{${}^1$Instituto de F\'isica, Pontificia Universidad Cat\'olica de Chile, Casilla 306, Santiago, Chile }

\begin{abstract} 
The yielding transition in amorphous materials, whether driven passively (simple shear) or actively, remains a fundamental open question in soft matter physics. While avalanche statistics at the critical point have been extensively studied, the emergence of the dynamic regime at yielding and the steady-state flow properties remain poorly understood. In particular, the significant variability observed in flow curves across different systems lacks a clear explanation. We determine, for the first time, the relationship between avalanche duration and size across the yielding transition, revealing how it evolves from quasistatic to dynamic flow regimes. This precise measurement is made using the Controlled Relaxation Time Model (CRTM), a new simulation framework that treats the relaxation time as a tunable parameter. CRTM reproduces known results in both limits and enables a direct analysis of the change of regime between them. Applying the model to different microscopic dynamics, we find that the existing scaling relation connecting critical exponents under flow holds for passive systems. However, active systems exhibit significant deviations, suggesting a missing ingredient in the current understanding of yielding.

\end{abstract}
\maketitle

\section{Introduction}
Foams, emulsions, grains, and suspensions are essential materials in various fields and industries, where controlling their mechanical and rheological properties is crucial~\cite{HILL2017,Wypych2022}. Despite their diverse microscopic constituents, these amorphous materials share a universal mechanical response: they remain mechanically stable at high packing densities but undergo a transition to a flowing state when the applied shear stress $\sigma$ exceeds a critical threshold~\cite{PhysRevE.68.011306,vanHecke_2010,PhysRevLett.99.178001,PhysRevLett.88.218301}. This yielding transition occurs at $\sigma_c$~\cite{RevModPhys.90.045006,RevModPhys.89.035005}, beyond which no mechanically stable state can withstand the applied stress, leading to a \emph{dynamic regime} where the material flows. Above $\sigma_c$, the rheology follows the Herschel-Bulkley law, $\dot{\gamma} \sim  (\sigma \!- \!\sigma_c)^\beta$~\cite{HB_original}, where $\dot{\gamma}$ is the strain rate and $\beta$ is the Herschel-Bulkley exponent  (sometimes reported as $n = 1/\beta$). This phenomenological law has been extensively validated through experiments and numerical simulations under simple shear (\emph{passive systems})~\cite{PhysRevLett.89.098303, PhysRevE.92.012305, PhysRevLett.90.068303, PhysRevE.82.031301}. Interestingly, comparable rheological behavior has also been reported in \emph{active systems}, including cell streaming in epithelial tissues and active glasses~\cite{Bi2015, PhysRevX.6.021011, PhysRevLett.131.188401, PhysRevE.106.L012601,ext_ac_in, CD1, 10.1039/c7sm01909b}. In these cases, once the active driving force exceeds a threshold $f_c$, jammed assemblies of self-propelled agents fluidize and exhibit a Herschel-Bulkley-like dependence of the strain rate on stress~\cite{ext_ac_in, CD1, D3SM01354E,Sharma2025, PhysRevE.84.040301, PhysRevLett.125.038003,D3SM00034F,10.1039/c7sm01909b}. Across both active and passive systems, the Herschel-Bulkley exponent exhibits significant variability, depending on factors such as dissipation mechanisms (e.g., inertial grains~\cite{PhysRevE.88.062206,Roy2015bubble}), dimensionality~\cite{PhysRevE.97.012603}, yielding event rules (e.g., mesoscopic elastoplastic models~\cite{C9SM01073D,Jagla2020,PhysRevLett.132.268203}), and more recently, different driving conditions (active/passive systems~\cite{CD1,PhysRevLett.131.188401}). Despite extensive studies, this variability remains largely unexplained, and a complete micro-mechanical description of the Herschel-Bulkley law is still missing.

\begin{figure*}[t]
\centering 
\includegraphics[width=0.975\textwidth]{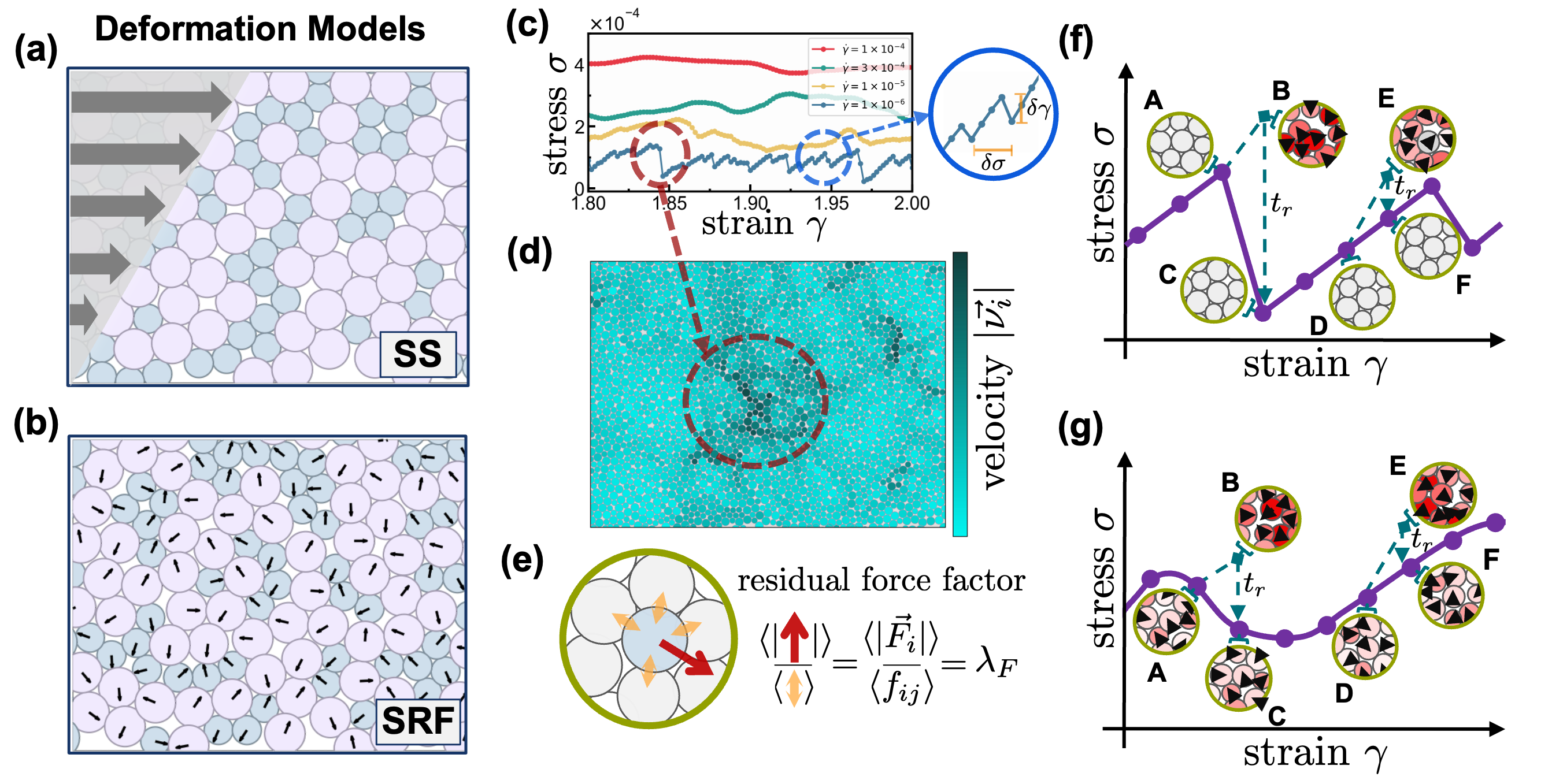}
\caption{\footnotesize \textbf{Deformation protocols and relaxation framework for yielding dynamics.} Schematic representation of the deformation scenarios: \textbf{(a)} Simple shear (SS), where the system is subjected to a velocity profile. \textbf{(b)} Self-random force (SRF), where each particle experiences a force $f$ applied along a fixed random direction $\hat{n}^{\mathrm{rnd}}_i$, which remains constant over time (infinite persistence). As in SS, when $f \!> \!f_c$, the system fails to reach mechanical equilibrium and continuously transitions between non-equilibrium states. \textbf{(c)} Stress–strain response obtained for different imposed strain rates $\dot{\gamma}$. At low $\dot{\gamma}$, the system exhibits intermittent stress drops associated with plastic avalanches. In this limit, close to the quasistatic regime, plastic events are triggered by a deformation increment $\delta\gamma$ and characterized by the resulting stress drop $\delta\sigma$, whereas higher rates lead to smoother flow curves. \textbf{(d)} Spatial map of the particle velocity magnitude $|\vec{\nu_i}|$ during one of these plastic events, showing a localized region of intense motion that propagates through the system, illustrating the avalanche dynamics. \textbf{(e)} Diagram illustrating the residual force factor $\lambda_F$, which quantifies the degree of mechanical equilibration as the ratio between the mean residual $\langle |\vec{F_i}| \rangle$, and contact force magnitudes ${\langle f_{ij} \rangle}$. Panels \textbf{(f)} and \textbf{(g)} display $\sigma$ \!vs.\! $\gamma$ curves under CRTM for shear deformation with relaxation times $t_r$ either sufficient or insufficient for the redistribution of stress to restore mechanical equilibrium. \textbf{(f)} $t_r$ sufficient: The athermal quasistatic limit is reached. The system starts in a mechanically equilibrated configuration (\textbf{A}, \textbf{D}). Each step applies an affine deformation, inducing interparticle forces and generating macroscopic shear stress $\sigma$ (\textbf{B}, \textbf{E}). Relaxation restores mechanical equilibrium (\textbf{C}, \textbf{F}), and the process repeats. \textbf{(g)} $t_r$ insufficient: Starting from near mechanical equilibrium (\textbf{A}, \textbf{D}), affine deformation (\textbf{B}, \textbf{E}) is followed by incomplete relaxation, leaving residual elastic forces (\textbf{C}, \textbf{F}) before the next deformation step.}
\label{fig:schemes_aqs_crtm} 
\end{figure*}

Understanding the yielding transition at a microscopic level requires resolving the dynamics of plastic events, which manifest as avalanches of rearrangements. At the yielding transition, successive plastic deformations organize into avalanches that can span the entire system~\cite{PhysRevLett.103.065501,PhysRevLett.93.016001}. The statistical properties of avalanche size distributions near yielding have been extensively studied using athermal quasistatic simulations (AQS) within both molecular dynamics and mesoscopic elastoplastic models~\cite{PhysRevE.74.016118, PhysRevE.79.066109, AQS5, PhysRevResearch.1.012002, PhysRevE.104.015002}. This quasistatic limit corresponds to very low strain rates, where plastic events appear as abrupt stress drops, as shown in Fig.~\ref{fig:schemes_aqs_crtm}.c.  Notably, these approaches yield consistent results across a wide range of conditions~\cite{D3SM01354E,PhysRevE.95.032902,Shang2020,PhysRevE.104.015002}. However, beyond $\sigma_c$, avalanche properties become highly system-dependent, as reflected in the significant variability of observed Herschel-Bulkley exponents. In mesoscopic elastoplastic models, the flow curve depends on the chosen relaxation scheme, and selecting an appropriate one  that ensures a physically meaningful avalanche relaxation remains a critical challenge~\cite{jocteur2024protocoldependenceavalanchesconstant,PhysRevE.97.012603,C9SM01073D}. In molecular dynamics simulations and experiments, avalanche statistics beyond the yielding point are far less understood and remain difficult to measure. The limited understanding of the connection between avalanche properties at the critical point and rheology at finite strain rates has hindered the development of a unified description of yielding. In particular, the relaxation properties of avalanches are believed to play a central role in determining the rheology above yielding~\cite{PhysRevE.97.012603}. The relationship between the characteristic relaxation time $T$ of an avalanche and its linear extension $l$, encoded in the dynamic exponent $z$ through $T \sim l^z$, has been pivotal in the description of depinning transitions~\cite{Xi2015}, a phenomenon closely related to yielding~\cite{Lin2014,PhysRevE.93.063005}. However, despite its importance, direct measurements of $z$ in the context of yielding transitions remain scarce and challenging~\cite{C9SM01073D,PhysRevE.97.012603,PhysRevE.103.042606,PhysRevLett.116.065501,PhysRevLett.127.108003}. Here we provide such a direct measurement within particle-based molecular dynamics, bridging the AQS and finite-rate limits.

We consider athermal soft disks in two dimensions, where particles interact in pairs through a potential $U(r_{ij})$, and examine two deformation schemes: simple shear (SS) and self-random force (SRF), the latter representing an active model with infinite persistence time. The particle dynamics in both cases follow the overdamped equations
\begin{equation}
    \vec{\nu}_i = \frac{d \vec{r}_i}{dt} = -D \frac{\partial U(r_{ij})}{\partial \vec{r}_i} + \vec{d c}_i,
\label{eq:overdamped_equation}
\end{equation}
where $\vec{r}_i$ and $\vec{\nu}_i$ denote the position and velocity of particle $i$, respectively, and $D$ is the overdamped constant. Under SS, the external driving term is given by $\vec{d c}_i = \dot{\gamma} (\vec{r}_i \cdot \hat{y}) \hat{x}$, which imposes a shear flow along $\hat{x}$ at strain rate $\dot{\gamma}$, with Lees--Edwards boundary conditions (Fig.~\ref{fig:schemes_aqs_crtm}.a). In the SRF protocol, each particle is subjected to a self-random force of fixed magnitude $f$, applied along a quenched random direction $\hat{n}^{\mathrm{\mathrm{rnd}}}_i$, resulting in $\vec{d c}_i = D f \hat{n}^{\mathrm{\mathrm{rnd}}}_i$ (Fig.~\ref{fig:schemes_aqs_crtm}.b). The rheology of these models has been previously studied \cite{CD1}, but no connection with the avalanche description in AQS models \cite{D3SM01354E} has been established. In this work, we present the first direct access to avalanche relaxation times and their scaling relation with avalanche spatial extent beyond AQS. To explore this gap, we introduce the \emph{Controlled Relaxation Time Model} (CRTM), which treats the relaxation time as a tunable parameter. Building on this microscopic foundation, this framework enables the computation of flow curves and spatial correlations, providing a direct route to test the fundamental scaling law relating strain rate and correlation length~\cite{Lin2014}. Interestingly, while this scaling law holds for passive systems, it breaks down in active ones, revealing the existence of a different microscopic mechanism that should lead to new directions for the study of the yielding transition.

\section{Results}
CRTM is built upon the well-established AQS method, a model that decouples affine deformation from system relaxation~\cite{Morse2021,PhysRevE.74.016118}. This approach is highly efficient in determining the static properties of avalanches at the critical stress, where the dynamics are sufficiently slow for the largest avalanches to fully relax~\cite{PhysRevE.74.016118, PhysRevE.104.015002, AQS5, PhysRevLett.93.016001}. Fig.~\ref{fig:schemes_aqs_crtm}.f provides a schematic representation of the AQS protocol. For SS, in the athermal quasistatic limit, each simulation step begins with an affine deformation of magnitude $\Delta \gamma$, which drives the system from a mechanically equilibrated state (\textbf{A} and \textbf{D}) to an out-of-equilibrium state (\textbf{B} and \textbf{E}), as illustrated in Fig.~\ref{fig:schemes_aqs_crtm}.f. This is followed by a relaxation process with a duration $t_r$ that exceeds the time needed for stress redistribution to bring the system to mechanical equilibrium. This ensures stability is reestablished at every step (points \textbf{C} and \textbf{F}), and the process is repeated until a total strain $\gamma$ is achieved. In our approach, $t_r$ is a tunable parameter that can be made smaller than the time required for full equilibration. As schematically illustrated in Fig.~\ref{fig:schemes_aqs_crtm}.g, when $t_r$ is insufficient for stress redistribution to restore equilibrium, relaxation remains incomplete, forcing the system to traverse out-of-equilibrium states. CRTM can be extended to the SRF by adopting the AQS implementation of this deformation scheme, where at each step the configuration is updated as $\vec{r}_i \rightarrow \vec{r}_i + \frac{L}{2\sqrt{N}}\,\Delta\gamma^{\mathrm{\mathrm{rnd}}}\,\hat{n}_i^{\mathrm{\mathrm{rnd}}} $~\cite{Morse2021,D3SM01354E}. A distinctive advantage of this method is its flexibility in the choice of relaxation dynamics at each step, which allows it to be tailored to the physical system under study. In this work, we primarily employ the \emph{Steepest Descent} method~\cite{SD}, since it corresponds to the direct integration of the overdamped dynamics of Eq.~\ref{eq:overdamped_equation} in the absence of driving. While the fidelity of discrete steepest-descent mappings in resolving individual inherent states has recently been debated, our analysis relies on macroscopic statistical observables rather than on this fine landscape structure~\cite{fg38-mtlf,Suryadevara2024basins}. Nevertheless, alternative procedures incorporating inertia (e.g., the FIRE algorithm~\cite{FIRE}) or nonlocal interactions can also be implemented. For reference, in the Supplementary Information we also present detailed simulations based on a \emph{Conjugate Gradient} relaxation scheme~\cite{CG-alg}, often employed in AQS simulations for numerical efficiency.

\begin{figure}[h]
\centering 
\includegraphics[width=0.49\textwidth]{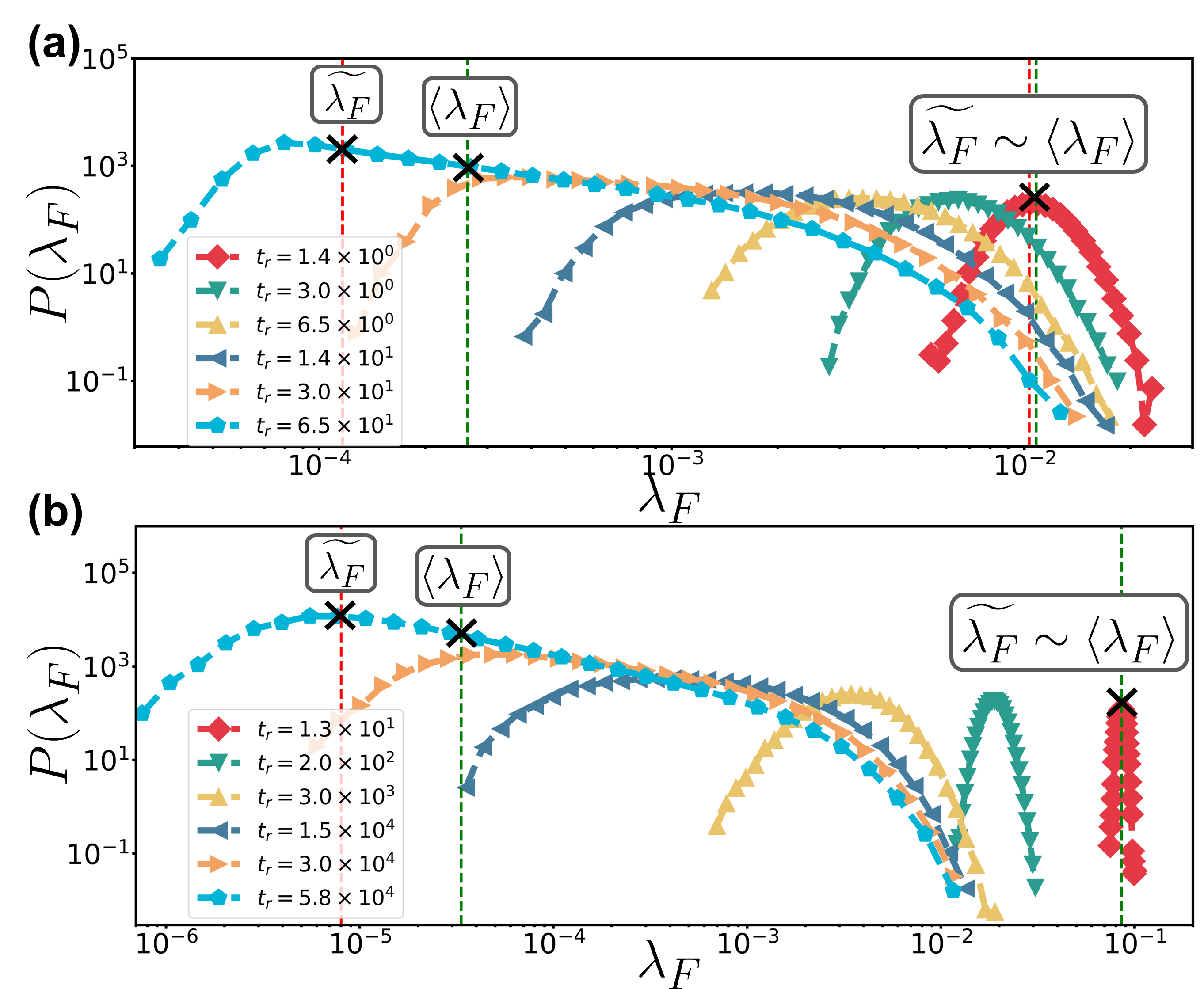}
\caption{\footnotesize \textbf{Crossover from quasistatic to dynamic behavior via residual force statistics.} Probability distributions $P(\lambda_F)$ for \textbf{(a)} SRF and \textbf{(b)} SS, shown for different relaxation times $t_r$. Vertical lines indicate, from left to right in each panel, the median $\tilde{\lambda_F}$ and mean $\langle \lambda_F \rangle$ for \textbf{(a)} $t_r = 6.5 \times 10 ^ 1$ and \textbf{(b)} $t_r = 5.8 \times 10 ^ 4$. The final pair of lines corresponds to the case where both values coincide, for \textbf{(a)} $t_r = 1.4 \times 10 ^ 0$ and \textbf{(b)} $t_r = 1.3 \times 10 ^ 1$. At low $t_r$, the distribution displays a single narrow peak at large $\lambda_F$ with $\tilde{\lambda_F} \!\approx\! \langle \lambda_F \rangle$. As $t_r$ increases, the peak shifts towards lower $\lambda_F$ and the distribution broadens, developing a long tail at large $\lambda_F$ associated with configurations still plastically active when $t_r$ is exhausted. In this transitional regime, the median and mean differ ($\tilde{\lambda_F} \!\neq\! \langle \lambda_F \rangle$). Simulations were performed with $N = 4096$ particles.}
\label{fig:lambdaDistributions_MLambdaf} 
\end{figure}

\begin{figure*}[t]
\centering 
\includegraphics[width=0.975\textwidth]{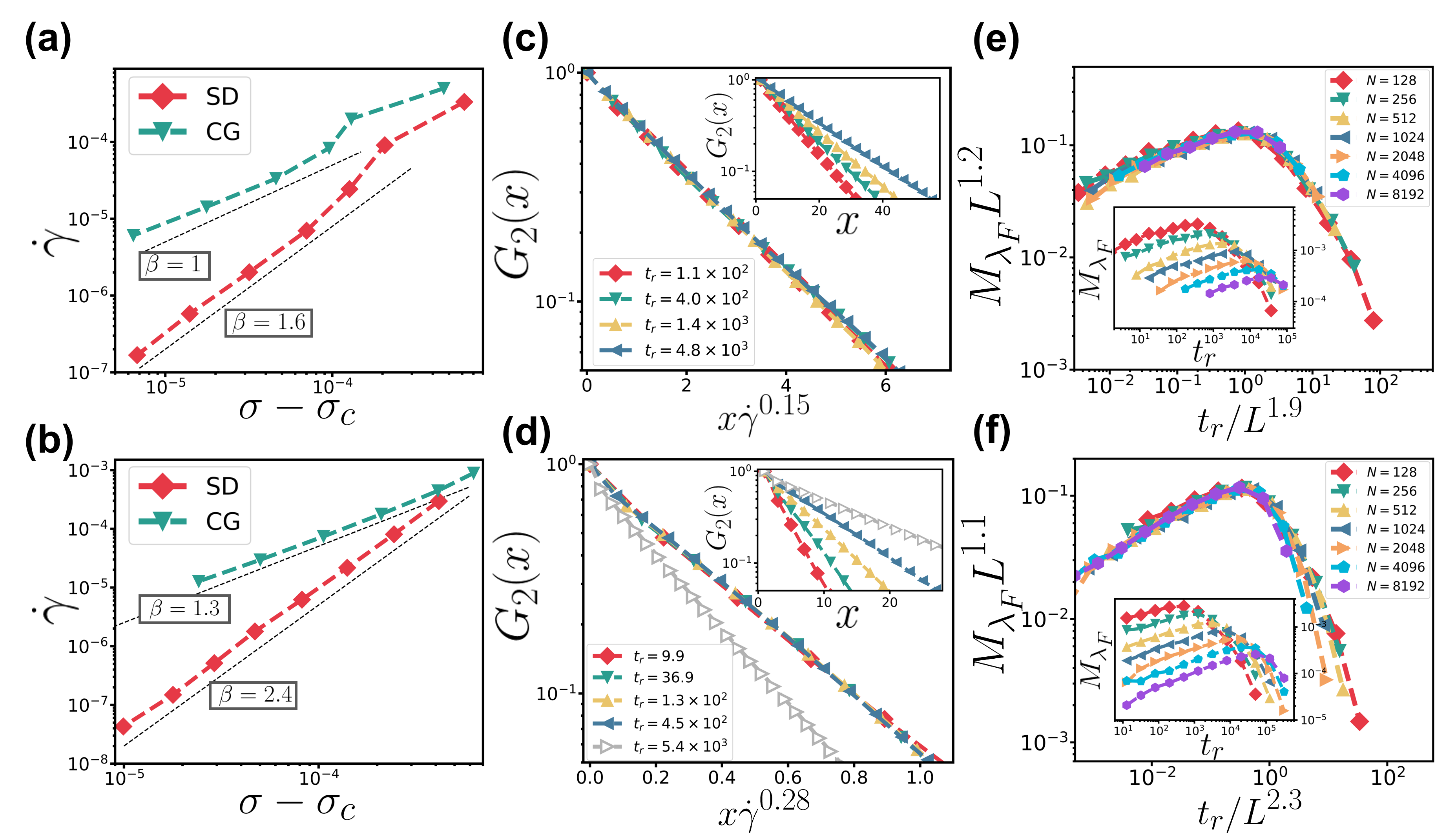}
\caption{\footnotesize \textbf{Flow curves, spatial correlations, and dynamic scaling across the yielding crossover.} Flow curves and spatial correlation functions obtained from CRTM simulations. Panels \textbf{(a)} and \textbf{(b)}: Flow curves for SRF and SS, respectively. Red markers correspond to the Steepest Descent relaxation method, which recovers the expected Herschel-Bulkley exponents~\cite{CD1}: $\beta = 1.6$ for SRF and $\beta = 2.4$ for SS. Green markers show results using the Conjugate Gradient method, yielding lower effective exponents: $\beta = 1.0$ for SRF and $\beta = 1.3$ for SS. Black lines represent best-fit power-law behaviors. Simulations were performed with $N = 4096$ particles. Panels \textbf{(c)} and \textbf{(d)}: Rescaled spatial correlation function $G_2(x)$ for SRF and SS, respectively, evaluated at various relaxation times $t_r$ using the Steepest Descent method. Insets display the unscaled data. The method yields $\nu / \beta$ values close to the expected ones~\cite{CD1}: $\nu / \beta = 0.15$ for SRF and $\nu / \beta = 0.28$ for SS. All reported $\nu / \beta$ values correspond to the best fit, with values within $\pm 0.03$ remaining within a reasonable range. In panel \textbf{(d)}, we include an additional curve with $t_r = 5.4 \times 10^3$, corresponding to a regime already entering the dynamic–quasi-static crossover. This is reflected in its inability to collapse with the other curves. Simulations were performed with $N = 8192$. Panels \textbf{(e)} and \textbf{(f)} show finite-size collapsed $M_{\lambda_F}$ as a function of relaxation time $t_r$ for SRF and SS respectively, using the Steepest Descent relaxation method. The insets display the raw data before rescaling. The extracted dynamic exponents are $z \!=\! 1.9$ for SRF and $z \!=\! 2.3$ for SS. All reported $z$ values correspond to the best fit, with values within $\pm 0.1$ remaining within a reasonable range.}
\label{fig:fluxCurves_and_spatialCorrelation} 
\end{figure*}

To quantify mechanical equilibrium, we introduce the \emph{residual force factor} (see schematic in Fig.~\ref{fig:schemes_aqs_crtm}.e), defined as
\begin{equation}
    \lambda_F = \frac{\langle |\vec{F_i}| \rangle}{\langle f_{ij} \rangle},
\label{eq:lambda_f}
\end{equation}
where $\langle |\vec{F_i}| \rangle$ is the mean magnitude of the residual force per particle and $\langle f_{ij} \rangle$ is the mean interparticle force. By construction, $\lambda_F \to 0$ indicates mechanical equilibrium, while finite values signal incomplete relaxation. In standard AQS simulations, equilibration is enforced by requiring $\lambda_F$ to fall below a numerical threshold, which in our simulations is set by machine precision, $\lambda_F\!\sim\!10^{-13}$. Within our CRTM framework, however, the relaxation may be interrupted before this condition is reached. 
Fig.~\ref{fig:lambdaDistributions_MLambdaf}.a,b shows the probability distribution $P(\lambda_F)$ for different $t_r$, using $\Delta\gamma\!= \!\Delta\gamma^{\mathrm{rnd}} \!=\! 8\!\times\! 10^{-4}$, under both SS and SRF driving. The simulations are similar to those used for AQS in Ref.~\cite{D3SM01354E}, but with a finite relaxation time $t_r$. Further details of the simulation setup, relaxation schemes, and the definition of $t_r$ are provided in the Methods and Supplementary Information. For large relaxation time $t_r$ the system explores progressively more stable configurations, approaching the athermal quasistatic limit. In this regime, a growing fraction of configurations ceases its plastic activity within $t_r$, populating the peak at low $\lambda_F$ observed in Fig.~\ref{fig:lambdaDistributions_MLambdaf}.a,b. The configurations still plastically active when $t_r$ is exhausted give rise to a long tail at large $\lambda_F$, which should vanish in the infinite relaxation time limit, leading to a single sharp peak at zero. Reaching this point with the Steepest Descent method is numerically expensive, and the tail persists for all accessible values of $t_r$. However, in the Supplementary Information we show that, by using the Conjugate Gradient relaxation method, the tail can be easily observed to be suppressed for sufficiently large $t_r$. For small $t_r$, the distribution of $\lambda_F$ reduces to a single narrow peak at large $\lambda_F$, whose position shifts as $t_r$ increases. In this regime, plastic activity is still ongoing in every configuration when the relaxation window ends, and thus the system remains out of equilibrium and represents a continuously flowing state. In the limit of small $t_r$, the rheology above yielding is expected to be recovered~\cite{CD1}. To test this, we define the strain rate as the imposed strain per relaxation time, $\dot{\gamma} = \frac{\Delta \gamma}{t_r}$, allowing a quantitative comparison between the steady-state statistics of the CRTM and the macroscopic flow curve. Fig.~\ref{fig:fluxCurves_and_spatialCorrelation}.a,b confirms that the expected rheology is indeed recovered for both SRF and SS driving~\cite{CD1}, including the phase separation observed in the active system. Consequently, in Fig.~\ref{fig:fluxCurves_and_spatialCorrelation}.a, we use the random strain rate defined as $\dot{\gamma} = \frac{\Delta \gamma^{\mathrm{\mathrm{rnd}}}}{t_r}$. The corresponding random stress is then given by $\sigma = \tfrac{\sqrt{N}}{2L} f - \frac{\Delta \gamma^{\mathrm{\mathrm{rnd}}}}{4D t_r}$, as previously defined in Ref.~\cite{CD1}. These quantities, which are directly related to particle velocities and driving forces, provide a natural basis for comparison with the SS. For comparison, in Fig.~\ref{fig:fluxCurves_and_spatialCorrelation}.a,b we also include results obtained using the Conjugate Gradient relaxation method (green). As expected, this numerical scheme yields different flow-curve exponents in both deformation protocols. Though it lacks a physical dynamical interpretation, these discrepancies highlight the strong dependence of the flow regime on the underlying microscopic relaxation dynamics. A detailed description of the implementation approaching the AQS limit of this method is provided in the Supplementary Information.

Another fundamental property of the yielding transition to test within the CRTM is the increasingly correlated dynamics as the system approaches the critical stress~\cite{PhysRevLett.103.065501,PhysRevResearch.1.012002,PhysRevLett.93.016001,PhysRevLett.108.178301,PhysRevLett.111.238301}. This growth in correlations, driven by avalanches that expand in both size and duration, naturally gives rise to a diverging length scale, $\xi \sim |\sigma - \sigma_c|^{-\nu}$, which governs the critical behavior. The spatial footprint of these events—visible as localized high-velocity patches in the velocity–magnitude field—is illustrated in Fig.~\ref{fig:schemes_aqs_crtm}.d. To quantify this length scale, we compute the spatial correlation function $G_2(x)$, derived from the variation of non-affine velocity, which provides a measure of avalanche dynamics~\cite{CD1}, details are presented in the Supplementary Information. These correlation curves collapse by plotting $G_2(x)$ against $x \dot{\gamma}^{\nu / \beta}$, allowing us to determine the $\nu / \beta$ exponent.  Fig.~\ref{fig:fluxCurves_and_spatialCorrelation}.c,d presents the collapsed correlations for both loading schemes. The results reproduce values consistent with prior studies~\cite{CD1}, confirming that CRTM effectively captures the flowing regime and the underlying critical behavior near yielding. \\

\begin{table*}[!ht ]
\setlength{\tabcolsep}{0pt}
\begin{tabular*}{\linewidth}{@{\extracolsep{\fill}}>{\itshape}wl{0.1\linewidth}wc{0.1\linewidth}wc{0.16\linewidth}wc{0.16\linewidth}wc{0.16\linewidth}wc{0.16\linewidth}}
\hline\hline
\theadl{\multirow{2}{*}{Exponent}} & \theadc{\multirow{2}{*}{Expression}} &  \multicolumn{2}{ c }{\small\itshape SRF (Active)} & \multicolumn{2}{ c }{\small\itshape SS (Passive)}\\
 & &  \theadc{Steepest Descent} & \theadc{Conjugate Gradient} & \theadc{Steepest Descent} & \theadc{Conjugate Gradient}\\
\hline
$\beta$        & $\dot{\gamma} \!\sim\! (\sigma - \sigma_c) ^\beta$   & 1.6           & 1             & 2.4           & 1.3 \\
$\nu / \beta$  & $\xi \!\sim\! \dot{\gamma}^{-\nu / \beta}$ & 0.15          & 0.32          & 0.28        & 0.4\\
$\delta$       &  $\langle \delta \sigma \rangle \!\sim\! L^{-\delta}$  & 1.14             & 1.14           & 1.04           & 1.04\\
$z$            & $T \sim l^z$  & 1.9 (5.5)           & 1.3 (2)           & 2.3 (2.5)           & 1.5 (1.4) \\
\hline\hline
\end{tabular*}
\caption{\footnotesize \textbf{Summary of critical exponents.} Results for SRF and SS loading schemes obtained using Steepest Descent and Conjugate Gradient relaxation methods. The exponents $\beta$ and $\nu / \beta$ are extracted from flow curve fits (Fig.~\ref{fig:fluxCurves_and_spatialCorrelation}.a,b) and spatial correlation analysis (Fig.~\ref{fig:fluxCurves_and_spatialCorrelation}.c,d), while $\delta$ values are taken from previous work~\cite{D3SM01354E}. The exponent $z$ is obtained by collapsing $M_{\lambda_F}$ curves (Fig.~\ref{fig:fluxCurves_and_spatialCorrelation}.e,f), while the values in parentheses correspond to $z$ computed from the scaling relation using the other exponents listed in this table. Details of the values obtained using the Conjugate Gradient method are provided in the Supplementary Information.}
\label{tab:compilation_exponents}
\end{table*}

Upon approaching the yielding transition—by reducing the strain rate or, equivalently, increasing the relaxation time $t_r$—the scaling properties break down as the correlation length reaches the system size (as observed in Fig.~\ref{fig:fluxCurves_and_spatialCorrelation}.d) and avalanches are expected to span the entire system. At this relaxation time, a crossover regime develops in which the distribution $P(\lambda_F)$ broadens as relaxed and plastically active configurations coexist, signaling the gradual transition from flowing to quasistatic behavior, as shown in Fig.~\ref{fig:lambdaDistributions_MLambdaf}.a,b. This coexistence defines the characteristic relaxation time scale $T$ at which avalanches reach the system size $L$~\cite{PhysRevE.74.016118,PhysRevResearch.1.012002}. Consequently, a finite-size scaling analysis can provide a direct estimate of the dynamic exponent $z$. To quantify this regime change, we compute the mean $\langle \lambda_F \rangle$ and the median $\tilde{\lambda_F}$. Their behavior is illustrated in Fig.~\ref{fig:lambdaDistributions_MLambdaf}.a,b, where the red and green vertical lines represent the median and mean, respectively. When a single peak dominates at low $t_r$, these values converge ($\tilde{\lambda_F} \!\approx\! \langle \lambda_F \rangle$), and the lines overlap. As $t_r$ increases, the distribution broadens and the growing population at low $\lambda_F$, together with the long tail at large $\lambda_F$, separates the mean from the median. To capture this separation, we define
\begin{equation}
M_{\lambda_F} \equiv \langle \lambda_F \rangle - \tilde{\lambda_F}.
\label{eq:factor_M}
\end{equation}
Fig.~\ref{fig:fluxCurves_and_spatialCorrelation}.e,f shows $M_{\lambda_F}$ as a function of the relaxation time $t_r$.$M_{\lambda_F}$ is small in both limits---at small $t_r$, where a single narrow peak dominates, and at very large $t_r$, where all configurations relax and both measures vanish---and grows as a power law across the crossover, reaching a maximum when $t_r$ becomes comparable to the relaxation time of system-spanning avalanches. This quantity provides a simple yet effective way to track the crossover from flowing to quasistatic behavior and to relate avalanche duration to system size. Assuming the scaling form $M_{\lambda_F}(t_r,L)=L^{-\alpha}\,\mathcal{F}(t_r/L^{z})$, which encodes how the correlation time scales with system size, $T\sim L^{z}$, we collapse the data in Fig.~\ref{fig:fluxCurves_and_spatialCorrelation}.e,f to obtain a direct estimate of the dynamic exponent $z$. At the finite-size crossover, where the correlation length becomes comparable to the system size, $\xi\simeq L$, this relation directly implies $T\sim \xi^{z}$. More generally, if the critical dynamics is governed by a single length scale, the same scaling relation is expected to hold for correlated regions of arbitrary size $\xi$.

The scaling relationship connecting the exponents that govern the yielding transition with avalanche statistics has been discussed in~\cite{Lin2014, PhysRevE.88.062206, C9SM01073D}. At the yielding transition, flow proceeds via successive avalanches of linear size up to the correlation length $\xi$, each releasing on average a shear stress $\langle \delta \sigma \rangle$ and occurring at a mean strain interval $\langle \delta \gamma \rangle$ (see Fig.~\ref{fig:schemes_aqs_crtm}.c). For the system to sustain flow, the time between successive avalanches, $\langle \delta \gamma \rangle / \dot{\gamma}$, must be of the order of the avalanche relaxation time $T$; otherwise, flow ceases. Assuming that avalanches at scales smaller than $\xi$ share the statistical properties of those at the critical stress, AQS results~\cite{Lin2014, D3SM01354E} give $\langle \delta \gamma \rangle \sim \langle \delta \sigma \rangle \sim \xi^{-\delta}$, where the system size $L$ is replaced by the correlation length $\xi$, and $\delta$ is obtained from the stress-drop distribution~\cite{D3SM01354E} (see table and Supplementary Information for details). Finally, using the scaling relation $T \sim \xi^{z}$, one obtains $\dot{\gamma}\sim \frac{\langle\delta \sigma \rangle}{T}\sim \xi^{-(\delta+z)}$ which leads to the scaling relation:
\begin{equation*}
    \frac{\nu}{\beta} = \frac{1}{\delta + z}.
    \label{eq:scaling_relation}
\end{equation*}

\noindent
The exponent $\delta$ is computed from $d_f$ and $\tau$, with values taken from our previous study~\cite{D3SM01354E}.

Table~\ref{tab:compilation_exponents} summarizes the full set of critical exponents, combining the results obtained here with $\delta$ from that work. In particular, the exponent $z$, extracted from the collapse of $M_{\lambda_F}$ using both the Steepest Descent and Conjugate Gradient methods, is compared with the value predicted by the scaling relation (shown in parentheses). This reveals a strong agreement between the collapsed exponent and the one predicted by the scaling relation for the passive system. Strikingly, no such agreement is observed for the active system, signaling a missing ingredient in the description of the yielding transition.

\section{Discussion}
This work addresses the yielding transition in amorphous materials from a new perspective, focused on the crossover between quasistatic and dynamic regimes. Although statistical analysis of avalanches in the quasistatic regime is a widely used tool to characterize plasticity, this approach alone does not allow for a complete description of the rheology above yielding. In this context, the CRTM framework is used to explore the residual force factor as an observable that enables us to identify whether the system operates under the quasistatic limit or under the dynamic one. The distribution of the residual force factor exhibits a robust regime change between the quasistatic and dynamic limits in both passive and active systems. A detailed study of these distributions shows a way to measure the relation between the duration of avalanches and their size. A finite-size scaling leads to a direct measurement of the $z$ exponent, a quantity that has long resisted precise characterization. In passive systems, this relationship generally holds; however, active systems display clear deviations whose origin remains to be identified in future work. Beyond the quasistatic limit, avalanche statistics remain poorly understood and difficult to measure, both in molecular dynamics simulations and in experiments. At finite strain rates, plastic events overlap in space and time, potentially modifying the quasistatic critical behavior~\cite{PhysRevLett.91.085702,PhysRevX.13.031034}; whether this overlap underlies these deviations remains an open question.

In addition, our results highlight the broader conceptual role and extensibility of the CRTM, which provides a continuous bridge between quasistatic and dynamic descriptions. This framework can be easily extended to systems with different types of interactions, including both passive and active cases, providing a simple way to explore avalanche dynamics across different relaxation methods without relying on separate algorithms. Overall, the CRTM offers a unified description of yielding regimes and a common ground for comparing different classes of amorphous systems.

\section{Methods}
The results presented in this work were obtained from athermal soft-disk simulations in two dimensions, performed under two loading schemes with a $1\!:\!1.4$ bidisperse mixture chosen to prevent crystallization~\cite{cristali_2}. For the SS protocol, Lees–Edwards boundary conditions were applied, whereas the SRF protocol employed periodic boundary conditions. Particles interacted through a purely repulsive Hertzian potential,
\begin{equation*}
U(r_{ij}) = \tfrac{2\epsilon}{5} \left(1 - r_{ij}/d_{ij}\right)^{5/2},
\end{equation*}
for $r_{ij} < d_{ij}$, and $U(r_{ij}) = 0$ otherwise, where $r_{ij}$ is the distance between particles $i$ and $j$, $d_{ij}$ the sum of their radii, and $\epsilon$ an energy scale. Unless otherwise noted, simulations were performed with strain increment $\Delta \gamma =\Delta \gamma^{\mathrm{rnd}} = 8 \times 10^{-4}$ at packing fraction $\phi = 0.95$, which ensures a wide window of driving rates in which the active model remains a homogeneous liquid, below the onset of motility-induced phase separation~\cite{CD1}. Further methodological details, together with robustness tests for different $\Delta \gamma$ values, are provided in the Supplementary Information.\\

\emph{Deformation schemes.--}  
The SS and SRF deformation models can be described in the dynamic regime by Eq.~\ref{eq:overdamped_equation}, and time is expressed in units of $t_0 = r_0^2 / (D \epsilon)$, where $r_0$ is the radius of the smallest particles (set to 1 in the bidisperse mixture). The term $\vec{dc}_i$ accounts for the driving contribution to the deformation. In the SRF model, each particle experiences a self-propulsion $\vec{dc}_i = D f \hat{n}_i^{\mathrm{\mathrm{rnd}}}$, where $f$ is the force magnitude and $\hat{n}_i^{\mathrm{\mathrm{rnd}}}$ a fixed random unit vector (see Fig.~\ref{fig:schemes_aqs_crtm}.b). In contrast, in SS the deformation is prescribed as $\vec{dc}_i = \dot{\gamma}(\vec{r}_i \cdot \hat{y}) \hat{x}$, producing a velocity profile consistent with shear deformation (see Fig.~\ref{fig:schemes_aqs_crtm}.a). To compare both cases, we define an equivalent strain rate for SRF, $\dot{\gamma}^{\mathrm{rnd}}$, based on the mean parallel velocity $\nu_{\parallel} = \tfrac{1}{N}\sum_{i=1}^N \vec{\nu}_i \cdot \hat{n}_i$, with $\hat{n}_i=\hat{x}$ in SS and $\hat{n}_i=\hat{n}_i^{\mathrm{rnd}}$ in SRF. In the passive case, Eq.~\ref{eq:overdamped_equation} together with the boundary conditions yields $\nu_{\parallel}^S = \dot{\gamma}L/(2\sqrt{N})$, from which we define $\dot{\gamma}^{\mathrm{rnd}} = (2\sqrt{N}/L)\,\nu_{\parallel}^{\mathrm{rnd}}$. The self-force can then be expressed as
$f = \tfrac{1}{N}\sum_{i=1}^N \big[ \tfrac{1}{D}\tfrac{d\vec{r}_i}{dt} + \tfrac{\partial U(r_{ij})}{\partial \vec{r}_i}\big]\cdot \hat{n}_i^{\mathrm{rnd}}$.  
Combining this with the imposed deformation, the SRF dynamics become

\begin{equation}
    \frac{d\vec{r}_i}{dt} = \frac{L}{2 \sqrt{N}} \dot{\gamma}^{\mathrm{rnd}} \hat{n}_i^{\mathrm{rnd}} 
    + D \left[ \overline{f}_{\parallel} \hat{n}_i^{\mathrm{rnd}} - \frac{\partial U(r_{ij})}{\partial \vec{r}_i} \right],
    \label{eq:overdamped_eq_SRF}
\end{equation}

\noindent
where $\overline{f}_{\parallel} = \tfrac{1}{N} \sum_{j=1}^{N} \tfrac{\partial U(r_{ij})}{\partial \vec{r}_j}\cdot \hat{n}_j^{\mathrm{rnd}}$ is the mean contact-force projection along the affine direction. Eq.~\ref{eq:overdamped_eq_SRF} represents the counterpart of the overdamped equation in Eq.~\ref{eq:overdamped_equation} for the SRF case, where the natural control parameter is $\dot{\gamma}^{\mathrm{rnd}}$ and the self-force $f$ is obtained as the response to this imposed deformation. This formulation enables the definition of an AQS algorithm for SRF~\cite{D3SM01354E}.

Analogous to the SS case, the stress in the SRF model is defined as
\begin{equation*}
\sigma^{\mathrm{rnd}} = \frac{1}{L^2}\,\frac{dU(r_{ij})}{d\gamma^{\mathrm{rnd}}}
             = \frac{1}{2L\sqrt{N}} \sum_{i=1}^N 
               \frac{\partial U(r_{ij})}{\partial \vec{r}_i}\cdot \hat{n}_i^{\mathrm{rnd}},
\end{equation*}
following Ref.~\cite{Morse2021}.  
Combining this definition with Eq.~\ref{eq:overdamped_eq_SRF} yields the compact expression
\begin{equation}
    \sigma^{\mathrm{rnd}} = \frac{\sqrt{N}}{2L} f - \frac{1}{4D}\,\dot{\gamma}^{\mathrm{rnd}}.
\end{equation}

Additional methodological details are provided in the Supplementary Information, which contains: a full account of the CRTM implementation for both SS and SRF protocols, including the step-by-step deformation rules and the definitions of relaxation time; a comparative analysis of the relaxation methods (Steepest Descent and Conjugate Gradient); the distributions of interparticle contact forces under SRF driving, resolved by strain rate; a characterization of the range of validity of the $\lambda_F$ statistics, bounded by the machine-precision floor; and a replication of our SRF results using an interaction potential that diverges at vanishing interparticle distance. These extended analyses ensure that the results reported in the main text are robust and not dependent on specific algorithmic or model choices.

\emph{Statistical Analysis.--} All reported quantities were obtained from ensembles of independent simulations. For each loading protocol (SS and SRF), we performed 10 realizations, unless otherwise stated, with distinct initial configurations drawn from the same distribution of particle sizes and packing fraction. After discarding a transient regime, measurements were taken in the steady state for each seed. Reported averages correspond to the mean over these realizations. Distributions were computed by pooling data from all realizations. No additional statistical tests or error bars were applied unless otherwise stated in figure legends.\\

\begin{acknowledgments}
\textbf{Acknowledgments.} We thank Edan Lerner and Ezequiel Ferrero for the fruitful discussions and comments on the manuscript. G.D. acknowledges funding from ANID FONDECYT No 1210656. C.V. acknowledges the support from ANID for Scholarship No. 21181971.GS. L.R. acknowledges the support from ANID for Scholarship No 21220738. 
\end{acknowledgments}

\section*{Data Availability}
The datasets generated and analyzed during this study are available at the following repository: \href{https://github.com/Poss9368/Bridging-the-Gap-Between-Avalanche-Relaxation-and-Yielding-Rheology}{GitHub repository} and will be archived at Zenodo (DOI to be provided upon acceptance). Additional data are available from the corresponding author upon reasonable request.

\section*{Code Availability}

The CRTM simulation code and analysis scripts used in this study are available at \href{https://github.com/Poss9368/Bridging-the-Gap-Between-Avalanche-Relaxation-and-Yielding-Rheology}{GitHub repository} and will be archived at Zenodo (DOI to be provided upon acceptance).

\section*{Correspondence}
Correspondence and requests for materials should be addressed to Leonardo Relmucao-Leiva (email: ltleiva@uc.cl).

\section*{Author Contributions}
C.V. conceived the study and developed the CRTM model and algorithms under the guidance of G.D. L.R.-L. performed most of the simulations and data analysis, with significant contributions from C.V. L.R.-L. and C.V. jointly wrote the manuscript. G.D. supervised the project and provided theoretical guidance. All authors discussed the results and approved the final version of the manuscript.

\section*{Competing Interests}
The authors declare no competing interests.

\bibliography{referencias}

@article{HILL2017,
title = {Foams: from nature to industry},
journal = {Advances in Colloid and Interface Science},
volume = {247},
pages = {496-513},
year = {2017},
note = {Dominique Langevin Festschrift: Four Decades Opening Gates in Colloid and Interface Science},
issn = {0001-8686},
doi = {https://doi.org/10.1016/j.cis.2017.05.013},
url = {https://www.sciencedirect.com/science/article/pii/S0001868616302998},
author = {Hill, C. and Eastoe, J.}
}

@incollection{Wypych2022,
title = {8 - Use in products},
editor = {Wypych, G.},
booktitle = {Handbook of Rheological Additives},
publisher = {ChemTec Publishing},
pages = {127-205},
year = {2022},
isbn = {978-1-927885-97-0},
doi = {https://doi.org/10.1016/B978-1-927885-97-0.50011-7},
url = {https://www.sciencedirect.com/science/article/pii/B9781927885970500117},
author = {Wypych, G.}
}

@article{RevModPhys.90.045006,
  title = {Deformation and flow of amorphous solids: Insights from elastoplastic models},
  author = {Nicolas, Alexandre and Ferrero, Ezequiel E. and Martens, Kirsten and Barrat, Jean-Louis},
  journal = {Rev. Mod. Phys.},
  volume = {90},
  issue = {4},
  pages = {045006},
  numpages = {63},
  year = {2018},
  month = {Dec},
  publisher = {American Physical Society},
  doi = {10.1103/RevModPhys.90.045006},
  url = {https://link.aps.org/doi/10.1103/RevModPhys.90.045006}
}

@article{PhysRevE.68.011306,
  title = {Jamming at zero temperature and zero applied stress: The epitome of disorder},
  author = {O'Hern, Corey S. and Silbert, Leonardo E. and Liu, Andrea J. and Nagel, Sidney R.},
  journal = {Phys. Rev. E},
  volume = {68},
  issue = {1},
  pages = {011306},
  numpages = {19},
  year = {2003},
  month = {Jul},
  publisher = {American Physical Society},
  doi = {10.1103/PhysRevE.68.011306},
  url = {https://link.aps.org/doi/10.1103/PhysRevE.68.011306}
}

@article{vanHecke_2010,
doi = {10.1088/0953-8984/22/3/033101},
url = {https://dx.doi.org/10.1088/0953-8984/22/3/033101},
year = {2009},
month = {dec},
publisher = {},
volume = {22},
number = {3},
pages = {033101},
author = {van Hecke, M.},
title = {Jamming of soft particles: geometry, mechanics, scaling and isostaticity},
journal = {Journal of Physics: Condensed Matter}
}

@article{PhysRevLett.99.178001,
  title = {Critical Scaling of Shear Viscosity at the Jamming Transition},
  author = {Olsson, Peter and Teitel, S.},
  journal = {Phys. Rev. Lett.},
  volume = {99},
  issue = {17},
  pages = {178001},
  numpages = {4},
  year = {2007},
  month = {Oct},
  publisher = {American Physical Society},
  doi = {10.1103/PhysRevLett.99.178001},
  url = {https://link.aps.org/doi/10.1103/PhysRevLett.99.178001}
}

@article{PhysRevLett.88.218301,
  title = {Coexistence of Liquid and Solid Phases in Flowing Soft-Glassy Materials},
  author = {Coussot, P. and Raynaud, J. S. and Bertrand, F. and Moucheront, P. and Guilbaud, J. P. and Huynh, H. T. and Jarny, S. and Lesueur, D.},
  journal = {Phys. Rev. Lett.},
  volume = {88},
  issue = {21},
  pages = {218301},
  numpages = {4},
  year = {2002},
  month = {May},
  publisher = {American Physical Society},
  doi = {10.1103/PhysRevLett.88.218301},
  url = {https://link.aps.org/doi/10.1103/PhysRevLett.88.218301}
}

@article{PhysRevLett.89.098303,
  title = {Shear-Induced Stress Relaxation in a Two-Dimensional Wet Foam},
  author = {Lauridsen, John and Twardos, Michael and Dennin, Michael},
  journal = {Phys. Rev. Lett.},
  volume = {89},
  issue = {9},
  pages = {098303},
  numpages = {4},
  year = {2002},
  month = {Aug},
  publisher = {American Physical Society},
  doi = {10.1103/PhysRevLett.89.098303},
  url = {https://link.aps.org/doi/10.1103/PhysRevLett.89.098303}
}

@article{PhysRevLett.103.065501,
  title = {Rate-Dependent Avalanche Size in Athermally Sheared Amorphous Solids},
  author = {Lema\^{\i}tre, A. and Caroli, C.},
  journal = {Phys. Rev. Lett.},
  volume = {103},
  issue = {6},
  pages = {065501},
  numpages = {4},
  year = {2009},
  month = {Aug},
  publisher = {American Physical Society},
  doi = {10.1103/PhysRevLett.103.065501},
  url = {https://link.aps.org/doi/10.1103/PhysRevLett.103.065501}
}

@article{PhysRevLett.93.016001,
  title = {Subextensive Scaling in the Athermal, Quasistatic Limit of Amorphous Matter in Plastic Shear Flow},
  author = {Maloney, C. and Lema\^{\i}tre, A.},
  journal = {Phys. Rev. Lett.},
  volume = {93},
  issue = {1},
  pages = {016001},
  numpages = {4},
  year = {2004},
  month = {Jul},
  publisher = {American Physical Society},
  doi = {10.1103/PhysRevLett.93.016001},
  url = {https://link.aps.org/doi/10.1103/PhysRevLett.93.016001}
}

@article{HB_original,
	author = {Herschel, Winslow H. and Bulkley, Ronald},
	date = {1926/08/01},
	doi = {10.1007/BF01432034},
	id = {Herschel1926},
	isbn = {1435-1536},
	journal = {Kolloid-Zeitschrift},
	number = {4},
	pages = {291--300},
	title = {Konsistenzmessungen von Gummi-Benzoll{\"o}sungen},
	url = {https://doi.org/10.1007/BF01432034},
	volume = {39},
	year = {1926}}

@article{PhysRevE.88.062206,
  title = {Effect of inertia on sheared disordered solids: Critical scaling of avalanches in two and three dimensions},
  author = {Salerno, K. Michael and Robbins, Mark O.},
  journal = {Phys. Rev. E},
  volume = {88},
  issue = {6},
  pages = {062206},
  numpages = {15},
  year = {2013},
  month = {Dec},
  publisher = {American Physical Society},
  doi = {10.1103/PhysRevE.88.062206},
  url = {https://link.aps.org/doi/10.1103/PhysRevE.88.062206}
}

@article{FIRE,
	author = {Bitzek, Erik and Koskinen, Pekka and G\"ahler, Franz and Moseler, Michael and Gumbsch, Peter},
	doi = {10.1103/PhysRevLett.97.170201},
	issue = {17},
	journal = {Phys. Rev. Lett.},
	month = {Oct},
	numpages = {4},
	pages = {170201},
	publisher = {American Physical Society},
	title = {Structural Relaxation Made Simple},
	url = {https://link.aps.org/doi/10.1103/PhysRevLett.97.170201},
	volume = {97},
	year = {2006}}

@article{PhysRevE.74.016118,
  title = {Amorphous systems in athermal, quasistatic shear},
  author = {Maloney, C. E. and Lema\^{\i}tre, A.},
  journal = {Phys. Rev. E},
  volume = {74},
  issue = {1},
  pages = {016118},
  numpages = {22},
  year = {2006},
  month = {Jul},
  publisher = {American Physical Society},
  doi = {10.1103/PhysRevE.74.016118},
  url = {https://link.aps.org/doi/10.1103/PhysRevE.74.016118}
}

@article{PhysRevResearch.1.012002,
  title = {Critical scaling for yield is independent of distance to isostaticity},
  author = {Thompson, Jacob D. and Clark, Abram H.},
  journal = {Phys. Rev. Res.},
  volume = {1},
  issue = {1},
  pages = {012002},
  numpages = {6},
  year = {2019},
  month = {Aug},
  publisher = {American Physical Society},
  doi = {10.1103/PhysRevResearch.1.012002},
  url = {https://link.aps.org/doi/10.1103/PhysRevResearch.1.012002}
}

@article{PhysRevE.79.066109,
  title = {Locality and nonlocality in elastoplastic responses of amorphous solids},
  author = {Lerner, Edan and Procaccia, Itamar},
  journal = {Phys. Rev. E},
  volume = {79},
  issue = {6},
  pages = {066109},
  numpages = {10},
  year = {2009},
  month = {Jun},
  publisher = {American Physical Society},
  doi = {10.1103/PhysRevE.79.066109},
  url = {https://link.aps.org/doi/10.1103/PhysRevE.79.066109}
}

@article{PhysRevE.95.032902,
  title = {Scaling of slip avalanches in sheared amorphous materials based on large-scale atomistic simulations},
  author = {Zhang, Dansong and Dahmen, Karin A. and Ostoja-Starzewski, Martin},
  journal = {Phys. Rev. E},
  volume = {95},
  issue = {3},
  pages = {032902},
  numpages = {12},
  year = {2017},
  month = {Mar},
  publisher = {American Physical Society},
  doi = {10.1103/PhysRevE.95.032902},
  url = {https://link.aps.org/doi/10.1103/PhysRevE.95.032902}
}

@article{Shang2020,
author = {Baoshuang Shang  and Pengfei Guan  and Jean-Louis Barrat },
title = {Elastic avalanches reveal marginal behavior in amorphous solids},
journal = {Proceedings of the National Academy of Sciences},
volume = {117},
number = {1},
pages = {86-92},
year = {2020},
doi = {10.1073/pnas.1915070117},
URL = {https://www.pnas.org/doi/abs/10.1073/pnas.1915070117},
eprint = {https://www.pnas.org/doi/pdf/10.1073/pnas.1915070117}}

@article{PhysRevE.104.015002,
  title = {Unified view of avalanche criticality in sheared glasses},
  author = {Oyama, Norihiro and Mizuno, Hideyuki and Ikeda, Atsushi},
  journal = {Phys. Rev. E},
  volume = {104},
  issue = {1},
  pages = {015002},
  numpages = {17},
  year = {2021},
  month = {Jul},
  publisher = {American Physical Society},
  doi = {10.1103/PhysRevE.104.015002},
  url = {https://link.aps.org/doi/10.1103/PhysRevE.104.015002}
}

@article{AQS5,
	author = {Ruscher, C{\'e}line and Rottler, J{\"o}rg},
	date = {2021/04/26},
	doi = {10.1007/s11249-021-01439-5},
	id = {Ruscher2021},
	isbn = {1573-2711},
	journal = {Tribology Letters},
	number = {2},
	pages = {64},
	title = {Avalanches in the Athermal Quasistatic Limit of Sheared Amorphous Solids: An Atomistic Perspective},
	url = {https://doi.org/10.1007/s11249-021-01439-5},
	volume = {69},
	year = {2021}}

@article{PhysRevLett.127.108003,
  title = {Instantaneous Normal Modes Reveal Structural Signatures for the Herschel-Bulkley Rheology in Sheared Glasses},
  author = {Oyama, Norihiro and Mizuno, Hideyuki and Ikeda, Atsushi},
  journal = {Phys. Rev. Lett.},
  volume = {127},
  issue = {10},
  pages = {108003},
  numpages = {6},
  year = {2021},
  month = {Sep},
  publisher = {American Physical Society},
  doi = {10.1103/PhysRevLett.127.108003},
  url = {https://link.aps.org/doi/10.1103/PhysRevLett.127.108003}
}

@article{Lin2014,
author = {Lin, J. and Lerner, E. and Rosso, A. and Wyart, M.},
title = {Scaling description of the yielding transition in soft amorphous solids at zero temperature},
journal = {Proceedings of the National Academy of Sciences},
volume = {111},
number = {40},
pages = {14382-14387},
year = {2014},
doi = {10.1073/pnas.1406391111},
URL = {https://www.pnas.org/doi/abs/10.1073/pnas.1406391111},
eprint = {https://www.pnas.org/doi/pdf/10.1073/pnas.1406391111}}

@Article{C9SM01073D,
author ={Ferrero, E. E. and Jagla, E. A.},
title  ={Criticality in elastoplastic models of amorphous solids with stress-dependent yielding rates},
journal  ={Soft Matter},
year  ={2019},
volume  ={15},
issue  ={44},
pages  ={9041-9055},
publisher  ={The Royal Society of Chemistry},
doi  ={10.1039/C9SM01073D},
url  ={http://dx.doi.org/10.1039/C9SM01073D}}

@article{ext_ac_in,
	author = {Mandal, Rituparno and Bhuyan, Pranab Jyoti and Chaudhuri, Pinaki and Dasgupta, Chandan and Rao, Madan},
	da = {2020/05/22},
	doi = {10.1038/s41467-020-16130-x},
	id = {Mandal2020},
	isbn = {2041-1723},
	journal = {Nature Communications},
	number = {1},
	pages = {2581},
	title = {Extreme active matter at high densities},
	ty = {JOUR},
	url = {https://doi.org/10.1038/s41467-020-16130-x},
	volume = {11},
	year = {2020}}

@article{PhysRevE.84.040301,
  title = {Active jamming: Self-propelled soft particles at high density},
  author = {Henkes, Silke and Fily, Yaouen and Marchetti, M. Cristina},
  journal = {Phys. Rev. E},
  volume = {84},
  issue = {4},
  pages = {040301},
  numpages = {4},
  year = {2011},
  month = {Oct},
  publisher = {American Physical Society},
  doi = {10.1103/PhysRevE.84.040301},
  url = {https://link.aps.org/doi/10.1103/PhysRevE.84.040301}
}

@article{PhysRevLett.125.038003,
  title = {Solid-Liquid Transition of Deformable and Overlapping Active Particles},
  author = {Loewe, Benjamin and Chiang, Michael and Marenduzzo, Davide and Marchetti, M. Cristina},
  journal = {Phys. Rev. Lett.},
  volume = {125},
  issue = {3},
  pages = {038003},
  numpages = {6},
  year = {2020},
  month = {Jul},
  publisher = {American Physical Society},
  doi = {10.1103/PhysRevLett.125.038003},
  url = {https://link.aps.org/doi/10.1103/PhysRevLett.125.038003}
}

@article{CD1,
	author = {Villarroel, C. and D{\"u}ring, G.},
	doi = {10.1039/D1SM00948F},
	issue = {43},
	journal = {Soft Matter},
	pages = {9944-9949},
	publisher = {The Royal Society of Chemistry},
	title = {Critical yielding rheology: from externally deformed glasses to active systems},
	url = {http://dx.doi.org/10.1039/D1SM00948F},
	volume = {17},
	year = {2021}}

@article{PhysRevLett.131.188401,
  title = {Random traction yielding transition in epithelial tissues},
  author = {Amiri, Aboutaleb and Duclut, Charlie and J\"ulicher, Frank and Popovi\ifmmode \acute{c}\else \'{c}\fi{}, Marko},
  journal = {Phys. Rev. Lett.},
  volume = {131},
  issue = {18},
  pages = {188401},
  numpages = {6},
  year = {2023},
  month = {Oct},
  publisher = {American Physical Society},
  doi = {10.1103/PhysRevLett.131.188401},
  url = {https://link.aps.org/doi/10.1103/PhysRevLett.131.188401}
}

@article{Morse2021,
author = {Peter K. Morse  and Sudeshna Roy  and Elisabeth Agoritsas  and Ethan Stanifer  and Eric I. Corwin  and M. Lisa Manning },
title = {A direct link between active matter and sheared granular systems},
journal = {Proceedings of the National Academy of Sciences},
volume = {118},
number = {18},
pages = {e2019909118},
year = {2021},
doi = {10.1073/pnas.2019909118},
URL = {https://www.pnas.org/doi/abs/10.1073/pnas.2019909118},
eprint = {https://www.pnas.org/doi/pdf/10.1073/pnas.2019909118}}

@article{D3SM00034F,
author ="Keta, Yann-Edwin and Mandal, Rituparno and Sollich, Peter and Jack, Robert L. and Berthier, Ludovic",
title  ="Intermittent relaxation and avalanches in extremely persistent active matter",
journal  ="Soft Matter",
year  ="2023",
volume  ="19",
issue  ="21",
pages  ="3871-3883",
publisher  ="The Royal Society of Chemistry",
doi  ="10.1039/D3SM00034F",
url  ="http://dx.doi.org/10.1039/D3SM00034F"}

@article{D3SM01354E,
author ={Villarroel, Carlos and Düring, Gustavo},
title  ={Avalanche properties at the yielding transition: from externally deformed glasses to active systems},
journal  ={Soft Matter},
year  ={2024},
volume  ={20},
issue  ={16},
pages  ={3520-3528},
publisher  ={The Royal Society of Chemistry},
doi  ={10.1039/D3SM01354E},
url  ={http://dx.doi.org/10.1039/D3SM01354E}}

@article{PhysRevE.97.012603,
  title = {Microscopic processes controlling the Herschel-Bulkley exponent},
  author = {Lin, Jie and Wyart, Matthieu},
  journal = {Phys. Rev. E},
  volume = {97},
  issue = {1},
  pages = {012603},
  numpages = {8},
  year = {2018},
  month = {Jan},
  publisher = {American Physical Society},
  doi = {10.1103/PhysRevE.97.012603},
  url = {https://link.aps.org/doi/10.1103/PhysRevE.97.012603}
}

@techreport{CG-alg,
	address = {USA},
	author = {Shewchuk, Jonathan R},
	institution = {Carnegie Mellon University},
	month = {March},
	publisher = {Carnegie Mellon University},
	title = {An introduction to the conjugate gradient method without the agonizing pain},
	year = {1994}}

@article{cristali_2,
	author = {Speedy,Robin J.},
	doi = {10.1063/1.478337},
	eprint = {https://doi.org/10.1063/1.478337},
	journal = {The Journal of Chemical Physics},
	number = {9},
	pages = {4559-4565},
	title = {Glass transition in hard disc mixtures},
	url = {https://doi.org/10.1063/1.478337},
	volume = {110},
	year = {1999}}

@incollection{SD,
	address = {San Diego},
	author = {Arora, J. S.},
	booktitle = {Introduction to Optimum Design (Second Edition)},
	doi = {https://doi.org/10.1016/B978-012064155-0/50008-2},
	edition = {Second Edition},
	editor = {Arora, J. S.},
	isbn = {978-0-12-064155-0},
	pages = {277-304},
	publisher = {Academic Press},
	title = {8 - Numerical methods for unconstrained optimum design},
	url = {https://www.sciencedirect.com/science/article/pii/B9780120641550500082},
	year = {2004}}

@article{PhysRevE.103.042606,
  title = {Criticality in sheared, disordered solids. II. Correlations in avalanche dynamics},
  author = {Clemmer, Joel T. and Salerno, K. Michael and Robbins, Mark O.},
  journal = {Phys. Rev. E},
  volume = {103},
  issue = {4},
  pages = {042606},
  numpages = {11},
  year = {2021},
  month = {Apr},
  publisher = {American Physical Society},
  doi = {10.1103/PhysRevE.103.042606},
  url = {https://link.aps.org/doi/10.1103/PhysRevE.103.042606}
}

@article{RevModPhys.89.035005,
  title = {Yield stress materials in soft condensed matter},
  author = {Bonn, Daniel and Denn, Morton M. and Berthier, Ludovic and Divoux, Thibaut and Manneville, S\'ebastien},
  journal = {Rev. Mod. Phys.},
  volume = {89},
  issue = {3},
  pages = {035005},
  numpages = {40},
  year = {2017},
  month = {Aug},
  publisher = {American Physical Society},
  doi = {10.1103/RevModPhys.89.035005},
  url = {https://link.aps.org/doi/10.1103/RevModPhys.89.035005}
}

@article{PhysRevE.92.012305,
  title = {Universal rescaling of flow curves for yield-stress fluids close to jamming},
  author = {Dinkgreve, M. and Paredes, J. and Michels, M. A. J. and Bonn, D.},
  journal = {Phys. Rev. E},
  volume = {92},
  issue = {1},
  pages = {012305},
  numpages = {17},
  year = {2015},
  month = {Jul},
  publisher = {American Physical Society},
  doi = {10.1103/PhysRevE.92.012305},
  url = {https://link.aps.org/doi/10.1103/PhysRevE.92.012305}
}

@article{PhysRevLett.90.068303,
  title = {Glassy Dynamics and Flow Properties of Soft Colloidal Pastes},
  author = {Cloitre, Michel and Borrega, R\'egis and Monti, Fabrice and Leibler, Ludwik},
  journal = {Phys. Rev. Lett.},
  volume = {90},
  issue = {6},
  pages = {068303},
  numpages = {4},
  year = {2003},
  month = {Feb},
  publisher = {American Physical Society},
  doi = {10.1103/PhysRevLett.90.068303},
  url = {https://link.aps.org/doi/10.1103/PhysRevLett.90.068303}
}

@article{PhysRevE.82.031301,
  title = {Statistical physics of elastoplastic steady states in amorphous solids: Finite temperatures and strain rates},
  author = {Karmakar, Smarajit and Lerner, Edan and Procaccia, Itamar and Zylberg, Jacques},
  journal = {Phys. Rev. E},
  volume = {82},
  issue = {3},
  pages = {031301},
  numpages = {11},
  year = {2010},
  month = {Sep},
  publisher = {American Physical Society},
  doi = {10.1103/PhysRevE.82.031301},
  url = {https://link.aps.org/doi/10.1103/PhysRevE.82.031301}
}

@article{Bi2015,
  author = {Bi, D. and Lopez, J. H. and Schwarz, J. M. and Manning, M. L.},
  title     = {A density-independent rigidity transition in biological tissues},
  journal   = {Nature Physics},
  year      = {2015},
  volume    = {11},
  number    = {12},
  pages     = {1074--1079},
  doi       = {10.1038/nphys3471},
  url       = {https://doi.org/10.1038/nphys3471},
  issn      = {1745-2481}
}

@article{PhysRevX.6.021011,
  title = {Motility-Driven Glass and Jamming Transitions in Biological Tissues},
  author = {Bi, Dapeng and Yang, Xingbo and Marchetti, M. Cristina and Manning, M. Lisa},
  journal = {Phys. Rev. X},
  volume = {6},
  issue = {2},
  pages = {021011},
  numpages = {13},
  year = {2016},
  month = {Apr},
  publisher = {American Physical Society},
  doi = {10.1103/PhysRevX.6.021011},
  url = {https://link.aps.org/doi/10.1103/PhysRevX.6.021011}
}

@article{PhysRevLett.108.178301,
  title = {Nonlocal Constitutive Relation for Steady Granular Flow},
  author = {Kamrin, Ken and Koval, Georg},
  journal = {Phys. Rev. Lett.},
  volume = {108},
  issue = {17},
  pages = {178301},
  numpages = {5},
  year = {2012},
  month = {Apr},
  publisher = {American Physical Society},
  doi = {10.1103/PhysRevLett.108.178301},
  url = {https://link.aps.org/doi/10.1103/PhysRevLett.108.178301}
}

@article{PhysRevLett.111.238301,
  title = {Nonlocal Rheology of Granular Flows across Yield Conditions},
  author = {Bouzid, Mehdi and Trulsson, Martin and Claudin, Philippe and Cl\'ement, Eric and Andreotti, Bruno},
  journal = {Phys. Rev. Lett.},
  volume = {111},
  issue = {23},
  pages = {238301},
  numpages = {5},
  year = {2013},
  month = {Dec},
  publisher = {American Physical Society},
  doi = {10.1103/PhysRevLett.111.238301},
  url = {https://link.aps.org/doi/10.1103/PhysRevLett.111.238301}
}

@misc{jocteur2024protocoldependenceavalanchesconstant,
      title={Protocol dependence for avalanches under constant stress in elastoplastic models}, 
      author = {Jocteur, T. and Bertin, E. and Mari, R. and Martens, K.},
      year={2024},
      eprint={2409.05444},
      archivePrefix={arXiv},
      primaryClass={cond-mat.soft},
      url={https://arxiv.org/abs/2409.05444}, 
}

@article{Jagla2020,
  title = {Tensorial description of the plasticity of amorphous composites},
  author = {Jagla, E. A.},
  journal = {Phys. Rev. E},
  volume = {101},
  issue = {4},
  pages = {043004},
  numpages = {10},
  year = {2020},
  month = {Apr},
  publisher = {American Physical Society},
  doi = {10.1103/PhysRevE.101.043004},
  url = {https://link.aps.org/doi/10.1103/PhysRevE.101.043004}
}

@article{PhysRevE.106.L012601,
  title = {Interplay between jamming and motility-induced phase separation in persistent self-propelling particles},
  author = {Yang, Jing and Ni, Ran and Ciamarra, Massimo Pica},
  journal = {Phys. Rev. E},
  volume = {106},
  issue = {1},
  pages = {L012601},
  numpages = {6},
  year = {2022},
  month = {Jul},
  publisher = {American Physical Society},
  doi = {10.1103/PhysRevE.106.L012601},
  url = {https://link.aps.org/doi/10.1103/PhysRevE.106.L012601}
}

@article{Sharma2025,
  author    = {Sharma, Rishabh and Karmakar, Smarajit},
  title     = {Activity-induced annealing leads to a ductile-to-brittle transition in amorphous solids},
  journal   = {Nature Physics},
  year      = {2025},
  volume    = {21},
  number    = {2},
  pages     = {253--261},
  doi       = {10.1038/s41567-024-02724-5},
  url       = {https://doi.org/10.1038/s41567-024-02724-5},
  issn      = {1745-2481},
  year      = {2025},
  month     = {Feb},
}

@article{PhysRevLett.132.268203,
  title = {Yielding Is an Absorbing Phase Transition with Vanishing Critical Fluctuations},
  author = {Jocteur, Tristan and Figueiredo, Shana and Martens, Kirsten and Bertin, Eric and Mari, Romain},
  journal = {Phys. Rev. Lett.},
  volume = {132},
  issue = {26},
  pages = {268203},
  numpages = {6},
  year = {2024},
  month = {Jun},
  publisher = {American Physical Society},
  doi = {10.1103/PhysRevLett.132.268203},
  url = {https://link.aps.org/doi/10.1103/PhysRevLett.132.268203}
}

@article{PhysRevE.93.063005,
  title = {From depinning transition to plastic yielding of amorphous media: A soft-modes perspective},
  author = {Tyukodi, Botond and Patinet, Sylvain and Roux, St\'ephane and Vandembroucq, Damien},
  journal = {Phys. Rev. E},
  volume = {93},
  issue = {6},
  pages = {063005},
  numpages = {12},
  year = {2016},
  month = {Jun},
  publisher = {American Physical Society},
  doi = {10.1103/PhysRevE.93.063005},
  url = {https://link.aps.org/doi/10.1103/PhysRevE.93.063005}
}

@article{Xi2015,
	author = {Xi, Bin and Luo, Meng-Bo and Vinokur, Valerii M. and Hu, Xiao},
	date = {2015/09/14},
	doi = {10.1038/srep14062},
	id = {Xi2015},
	isbn = {2045-2322},
	journal = {Scientific Reports},
	number = {1},
	pages = {14062},
	title = {Depinning Transition of a Domain Wall in Ferromagnetic Films},
	url = {https://doi.org/10.1038/srep14062},
	volume = {5},
	year = {2015}}

@article{PhysRevLett.116.065501,
  title = {Driving Rate Dependence of Avalanche Statistics and Shapes at the Yielding Transition},
  author = {Liu, Chen and Ferrero, Ezequiel E. and Puosi, Francesco and Barrat, Jean-Louis and Martens, Kirsten},
  journal = {Phys. Rev. Lett.},
  volume = {116},
  issue = {6},
  pages = {065501},
  numpages = {5},
  year = {2016},
  month = {Feb},
  publisher = {American Physical Society},
  doi = {10.1103/PhysRevLett.116.065501},
  url = {https://link.aps.org/doi/10.1103/PhysRevLett.116.065501}
}

@article{10.1039/c7sm01909b,
    author = {Liao, Qinyi and Xu, Ning},
    title = {Criticality of the zero-temperature jamming transition probed by self-propelled particles},
    journal = {Soft Matter},
    volume = {14},
    number = {5},
    pages = {853-860},
    year = {2018},
    month = {02},
    issn = {1744-683X},
    doi = {10.1039/c7sm01909b},
    url = {https://doi.org/10.1039/c7sm01909b},
    eprint = {https://pubs.rsc.org/sm/article-pdf/14/5/853/5796920/c7sm01909b.pdf},
}

@article{fg38-mtlf,
  title = {Numerically discovered inherent states are always protocol dependent in jammed packings},
  author = {Bautista, Eddie and Corwin, Eric I.},
  journal = {Phys. Rev. E},
  volume = {113},
  issue = {4},
  pages = {045408},
  numpages = {6},
  year = {2026},
  month = {Apr},
  publisher = {American Physical Society},
  doi = {10.1103/fg38-mtlf},
  url = {https://link.aps.org/doi/10.1103/fg38-mtlf}
}

@misc{Suryadevara2024basins,
  title        = {The Basins of Attraction of Soft Sphere Packings Are Not Fractal},
  author       = {Suryadevara, Praharsh and Casiulis, Mathias and Martiniani, Stefano},
  year         = {2024},
  eprint       = {2409.12113},
  archivePrefix= {arXiv},
  primaryClass = {cond-mat.soft},
  note         = {arXiv:2409.12113}
}

@article{Roy2015bubble,
  title = {Rheology, diffusion, and velocity correlations in the bubble model},
  author = {Roy, Arka Prabha and Karimi, Kamran and Maloney, Craig E.},
  journal = {arXiv preprint arXiv:1508.00810},
  year = {2015},
  eprint = {1508.00810},
  archivePrefix = {arXiv},
  primaryClass = {cond-mat.soft}
}

@article{PhysRevX.13.031034,
  title = {Scaling Description of Dynamical Heterogeneity and Avalanches of Relaxation in Glass-Forming Liquids},
  author = {Tahaei, Ali and Biroli, Giulio and Ozawa, Misaki and Popovi\ifmmode \acute{c}\else \'{c}\fi{}, Marko and Wyart, Matthieu},
  journal = {Phys. Rev. X},
  volume = {13},
  issue = {3},
  pages = {031034},
  numpages = {20},
  year = {2023},
  month = {Sep},
  publisher = {American Physical Society},
  doi = {10.1103/PhysRevX.13.031034},
  url = {https://link.aps.org/doi/10.1103/PhysRevX.13.031034}
}

@article{PhysRevLett.91.085702,
  title = {Driving Rate Effects on Crackling Noise},
  author = {White, Robert A. and Dahmen, Karin A.},
  journal = {Phys. Rev. Lett.},
  volume = {91},
  issue = {8},
  pages = {085702},
  numpages = {4},
  year = {2003},
  month = {Aug},
  publisher = {American Physical Society},
  doi = {10.1103/PhysRevLett.91.085702},
  url = {https://link.aps.org/doi/10.1103/PhysRevLett.91.085702}
}
\end{document}


\title{Supplementary Information:\\ Bridging the Gap Between Avalanche Relaxation and Yielding Rheology}

\author{Leonardo Relmucao-Leiva${}^{1}$, Carlos Villarroel${}^{1}$ and  Gustavo D\"uring${}^{1}$  }
\affiliation{${}^1$Instituto de F\'isica, Pontificia Universidad Cat\'olica de Chile, Casilla 306, Santiago, Chile }
\maketitle

\appendix
\section{CRTM detail for SS and SRF}
\label{app:model_details}
\emph{AQS algorithm}.--  CRTM is derived from the standard AQS algorithm~\cite{D3SM01354E}. Explaining it in both the passive and active cases helps clarify the behavior of the model. Most quasistatic simulations are based on the idea of decoupling the time scale on which the affine deformation is applied from the time scale of the relaxation process that takes place in response to that deformation.

\begin{figure}[h]
\centering 
\includegraphics[width=0.5\textwidth]{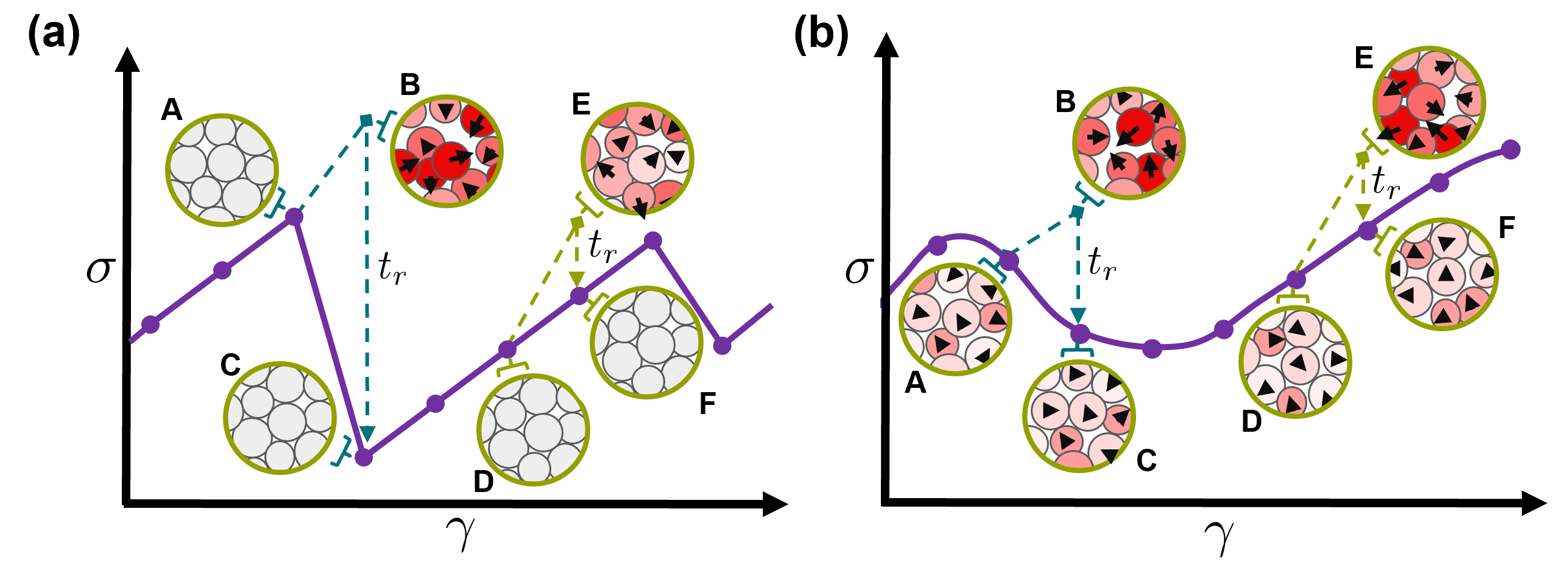}
\caption{\footnotesize\textbf{CRTM deformation and relaxation.} Panels \textbf{(a)} and \textbf{(b)} show schematic $\sigma$--$\gamma$ curves obtained with CRTM for relaxation times $t_r$ that are either sufficient or insufficient to restore mechanical equilibrium after each deformation step. \textbf{(a)} When $t_r$ is sufficiently large, the system reaches the athermal quasistatic limit: starting from an equilibrated configuration (\textbf{A}, \textbf{D}), an affine deformation induces an increase in stress (\textbf{B}, \textbf{E}), after which the system fully relaxes back to mechanical equilibrium (\textbf{C}, \textbf{F}). \textbf{(b)} When $t_r$ is too small, the affine deformation (\textbf{B}, \textbf{E}) is followed by only partial relaxation, leaving finite residual forces in configurations (\textbf{C}, \textbf{F}) before the next step. This incomplete relaxation produces flowing states characteristic of the dynamic regime.}
\label{fig:esquema_CRTM}
\end{figure}

To illustrate this, we start from the overdamped dynamics under simple shear (SS),
\begin{equation}
    \frac{d\vec{r}_i}{dt} = \dot{\gamma} (\vec{r}_i \cdot \hat{y}) \hat{x}
    - D \frac{\partial U(r_{ij})}{\partial \vec{r}_i},
\end{equation}
where the first term represents the imposed affine shear flow and the second term accounts for the relaxation driven by elastic forces. In the quasistatic limit, the strain rate $\dot{\gamma}$ is taken to zero while keeping a finite strain increment $d\gamma$ between successive configurations, and the system is allowed to fully relax back to mechanical equilibrium after each increment. Formally, the displacement of particle $i$ associated with a small strain increment $d\gamma$ can be written as
\begin{equation}
    d\vec{r}_i = (\vec{r}_i \cdot \hat{y}) \hat{x}\, d\gamma
    - D \int_0^{\infty} \frac{\partial U(r_{ij})}{\partial \vec{r}_i}\, dt,
    \label{eq:integrated_overdamped_eq_SS}
\end{equation}
where the first term corresponds to the instantaneous affine deformation and the second term encodes the subsequent relaxation towards mechanical equilibrium. In practice, the AQS protocol imposes a finite affine strain increment $\Delta \gamma$ at each step. For SS, the particle positions are updated according to
\begin{equation}
    \vec{r}_i \;\rightarrow\; \vec{r}_i + \Delta \gamma\, (\vec{r}_i \cdot \hat{y}) \hat{x}.
    \label{eq:aqs_SS}
\end{equation}
This deformation step is illustrated in Fig.~\ref{fig:esquema_CRTM}.a, where the configuration evolves from \textbf{A} and \textbf{D} to \textbf{B} and \textbf{E}. After the affine deformation is applied, the system is relaxed by minimizing its potential energy until mechanical equilibrium is recovered, i.e., until the residual forces fall below a prescribed threshold. This relaxation step is represented by the transition from \textbf{B} and \textbf{E} to the relaxed states \textbf{C} and \textbf{F}.

To derive the AQS formulation for SRF, we follow the same procedure as in the passive case. We begin with the overdamped dynamics in the flowing regime,
\begin{equation}
    \frac{d\vec{r}_i}{dt} =
    \frac{L}{2 \sqrt{N}}\,\dot{\gamma}^{\mathrm{rnd}}\,\hat{n}_i^{\mathrm{rnd}}
    + D\left[\,\overline{f}_{\parallel}\,\hat{n}_i^{\mathrm{rnd}}
    - \frac{\partial U}{\partial \vec{r}_i}\right],
    \label{eq:overdamped_eq_SRF}
\end{equation}
where the first term represents the imposed active driving along the fixed random direction
$\hat{n}_i^{\mathrm{rnd}}$, and the second term accounts for the relaxation driven by elastic forces and the mean projection of contact forces.\\

As in SS, the quasistatic limit corresponds to taking $\dot{\gamma}^{\mathrm{rnd}} \to 0$
while keeping a finite increment $d\gamma^{\mathrm{rnd}}$ between successive configurations.
In this limit, the system has enough time to relax fully after each deformation step, and
Eq.~\eqref{eq:overdamped_eq_SRF} becomes time–independent. Integrating over time yields
\begin{equation}
    d\vec{r}_i =
    \frac{L}{2\sqrt{N}}\, d\gamma^{\mathrm{rnd}}\,\hat{n}_i^{\mathrm{rnd}}
    + D\int_0^{\infty}
    \left[\,\overline{f}_{\parallel}\,\hat{n}_i^{\mathrm{rnd}}
           - \frac{\partial U}{\partial \vec{r}_i}\right] dt,
    \label{eq:integrated_overdamped_eq_SRF}
\end{equation}
The first term represents the imposed
active deformation, analogous to the affine shear term in SS, while the second term encodes
the complete relaxation towards mechanical equilibrium.

Numerically, the condition $t\to\infty$ is implemented by allowing the system to relax until
mechanical balance is reached. In the presence of active driving, this balance requires
$\frac{\partial U}{\partial \vec{r}_i} = \overline{f}_{\parallel}\,\hat{n}_i^{\mathrm{rnd}},$
ensuring that the residual forces vanish in the direction of self–propulsion. Thus, the AQS
algorithm for SRF follows the same two–step structure as in SS: an imposed deformation
controlled by $d\gamma^{\mathrm{rnd}}$ followed by a full relaxation to the corresponding
equilibrium configuration.\\

\emph{CRTM algorithm.--}  
The CRTM extends the AQS procedure by controlling the amount of relaxation allowed after each deformation step. Instead of relaxing the system until full mechanical equilibrium is reached, the relaxation is truncated at a prescribed time $t_r$. By tuning $t_r$, CRTM generalizes the AQS protocol, recovering the quasistatic limit when $t_r \to \infty$ and accessing flowing states when relaxation is restricted.

For SS, the update rule follows directly from the integrated overdamped dynamics in Eq.~\eqref{eq:integrated_overdamped_eq_SS}. After applying the affine increment $\Delta\gamma$, the elastic forces are integrated only up to time $t_r$, yielding
\begin{equation}
    \vec{r}_i \;\rightarrow\;
    \vec{r}_i
    + \Delta\gamma\,(\vec{r}_i \cdot \hat{y})\hat{x}
    - D \int_0^{t_r} \frac{\partial U}{\partial \vec{r}_i}\, dt .
    \label{eq:desp_SS}
\end{equation}

The SRF update rule is obtained analogously from the quasistatic SRF expression in Eq.~\eqref{eq:integrated_overdamped_eq_SRF}, replacing the full relaxation by an integration over the finite time interval $[0,t_r]$:
\begin{equation}
    \vec{r}_i \;\rightarrow\;
    \vec{r}_i
    + \frac{L}{2\sqrt{N}}\, \Delta\gamma^{\mathrm{rnd}}\,\hat{n}_i^{\mathrm{rnd}}
    + D \int_0^{t_r}
        \left[\overline{f}_{\parallel}\,\hat{n}_i^{\mathrm{rnd}}
              - \frac{\partial U}{\partial \vec{r}_i}
        \right] dt .
    \label{eq:desp_SRF_CRTM}
\end{equation}

In both loading schemes, the CRTM update consists of two steps:  
(i) the imposed deformation, taking the system from configurations \textbf{A} and \textbf{D} to \textbf{B} and \textbf{E}; and  
(ii) a relaxation stage of duration $t_r$, which moves the system toward \textbf{C} and \textbf{F} as shown in Fig.~\ref{fig:esquema_CRTM}.b.  
For sufficiently large $t_r$, the full quasistatic behavior is recovered, while smaller values of $t_r$ allow the system to explore states with finite residual forces, capturing dynamical effects beyond the AQS limit.

\section{Relaxation Methods}
\label{app:relax_methods}
In practice, the terms on the right-hand side of Eq.~\ref{eq:desp_SS} and Eq.~\ref{eq:desp_SRF_CRTM} are computed through an energy relaxation process, which can be adjusted as needed. After the deformation step, we employ two distinct relaxation methods. The Steepest Descent~\cite{SD} (SD) method implements a relaxation dynamics consistent with the overdamped equations of motion and therefore tends to be computationally expensive. The Conjugate Gradient~\cite{CG-alg} (CG) method offers a more efficient numerical alternative. Moreover, CG enables meaningful comparisons by providing an alternative way to analyze the system’s dynamics.

To ensure a consistent definition of the relaxation time $t_r$ in both methods, we express the total displacement as:

\begin{equation*} 
\vec{r}_{t_r, i} - \vec{r}_{0, i} = \sum_j^{Ns} \vec{\nu}_{i,j} \Delta t,
\label{eq:origin_delta_t} \end{equation*}

\noindent 
where $\Delta t = t_r / M$, and $Ns$ is the number of discretization steps. In the SD method, the velocity at each step is given by ${\vec{\nu}_{i,j} = \vec{F}_{i,j} / D}$, where $\vec{F}_{i,j}$ represents the elastic forces acting on the $i$-th particle. In the CG method, the velocity update rule is modified as ${\vec{\nu}_{i,j} \to \vec{\nu}_{i,j} + \vec{\alpha}}$, where $\vec{\alpha}$ optimizes the minimization process by improving convergence.

\section{Numerical Limits of $\lambda_F$}
\label{app:numerical_limits}

\begin{figure}[h]
\centering 
\includegraphics[width=0.48\textwidth]{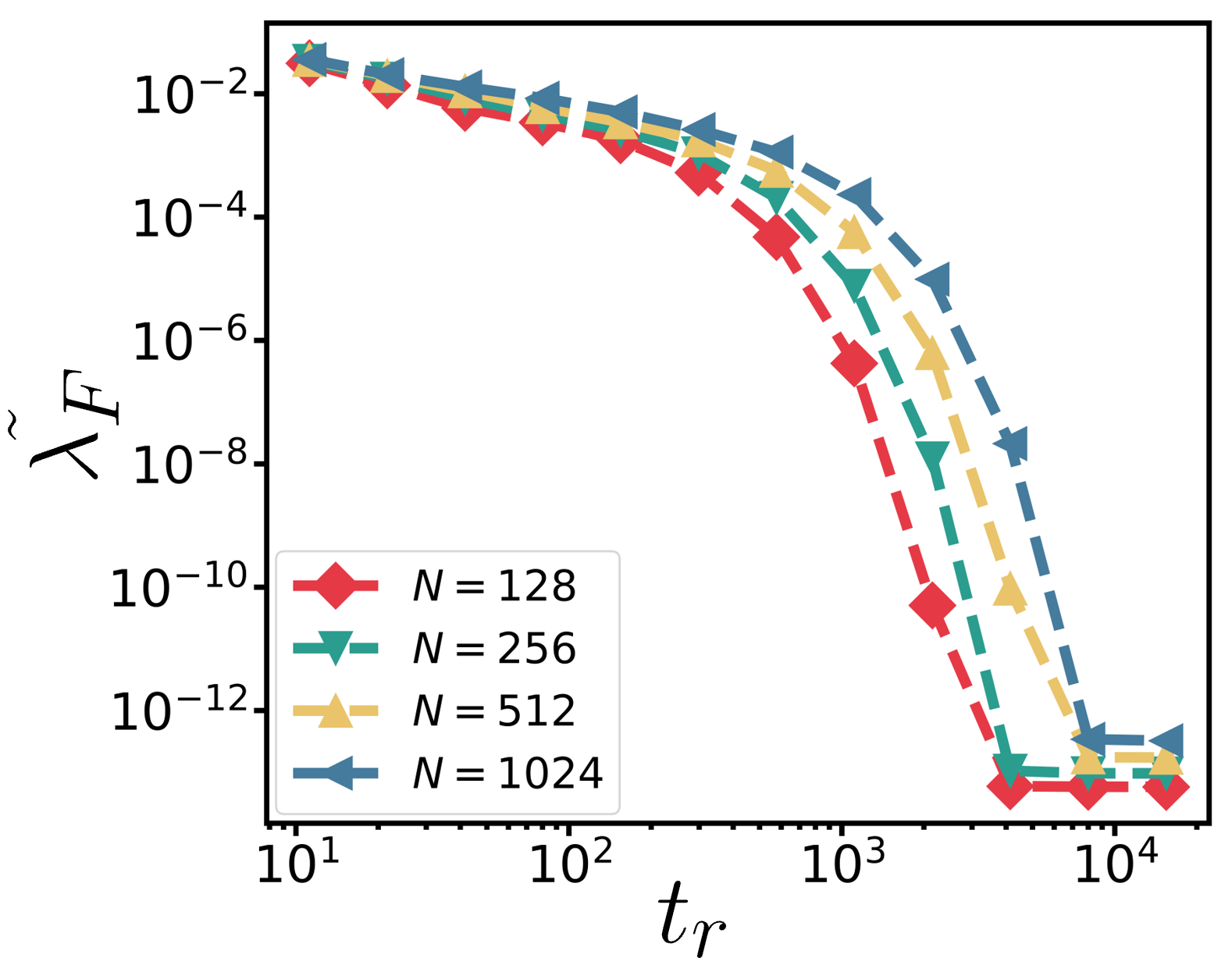}
\caption{\footnotesize \textbf{Numerical precision floor of the residual force factor.} Median of $\lambda_F$ as a function of the relaxation time $t_r$ for different system sizes $N$, using the Conjugate Gradient relaxation method under SS driving. The median decreases with $t_r$ until saturating at a floor set by machine precision ($\lambda_F\!\sim\!10^{-13}$), which shifts to larger $t_r$ with increasing system size. All scaling analyses are restricted to the regime where the median remains well above this floor. }
\label{fig:median_lambdaf_floor}
\end{figure}

To identify the numerical resolution limit of the residual force factor, we analyze the evolution of the median of $\lambda_F$ as a function of the relaxation time. Fig.~\ref{fig:median_lambdaf_floor} shows the median obtained for different system sizes using the Conjugate Gradient relaxation method under SS driving. 

The median decreases continuously as the relaxation time increases. For sufficiently large $t_r$, the curves approach a saturation value near $\lambda_F \sim 10^{-13}$, corresponding to the numerical precision floor. This behavior sets the accessible range of $\lambda_F$ before numerical effects dominate the measurements. All scaling analyses reported in this work are restricted to the regime where the characteristic values of $\lambda_F$ remain well above this floor.

\newpage
\section{Approach to the Fully Elastic Regime at Long Relaxation Times}
\label{app:Prob_tr_grande}
For sufficiently long relaxation times, every configuration is expected to reach the regime in which relaxation proceeds purely elastically, with the residual force driven toward zero as the system settles into a local energy minimum. Reaching this limit in practice is demanding: it requires both large $t_r$ and an efficient relaxation method. Additionally, before all configurations attain a fully relaxed state, the smallest residual forces already reach the machine-precision floor, making an accumulation of $P(\lambda_F)$ at this floor unavoidable. Even so, the Conjugate Gradient method under SS driving makes this approach to the fully elastic regime clearly visible.

\begin{figure}[h]
\centering 
\includegraphics[width=0.45\textwidth]{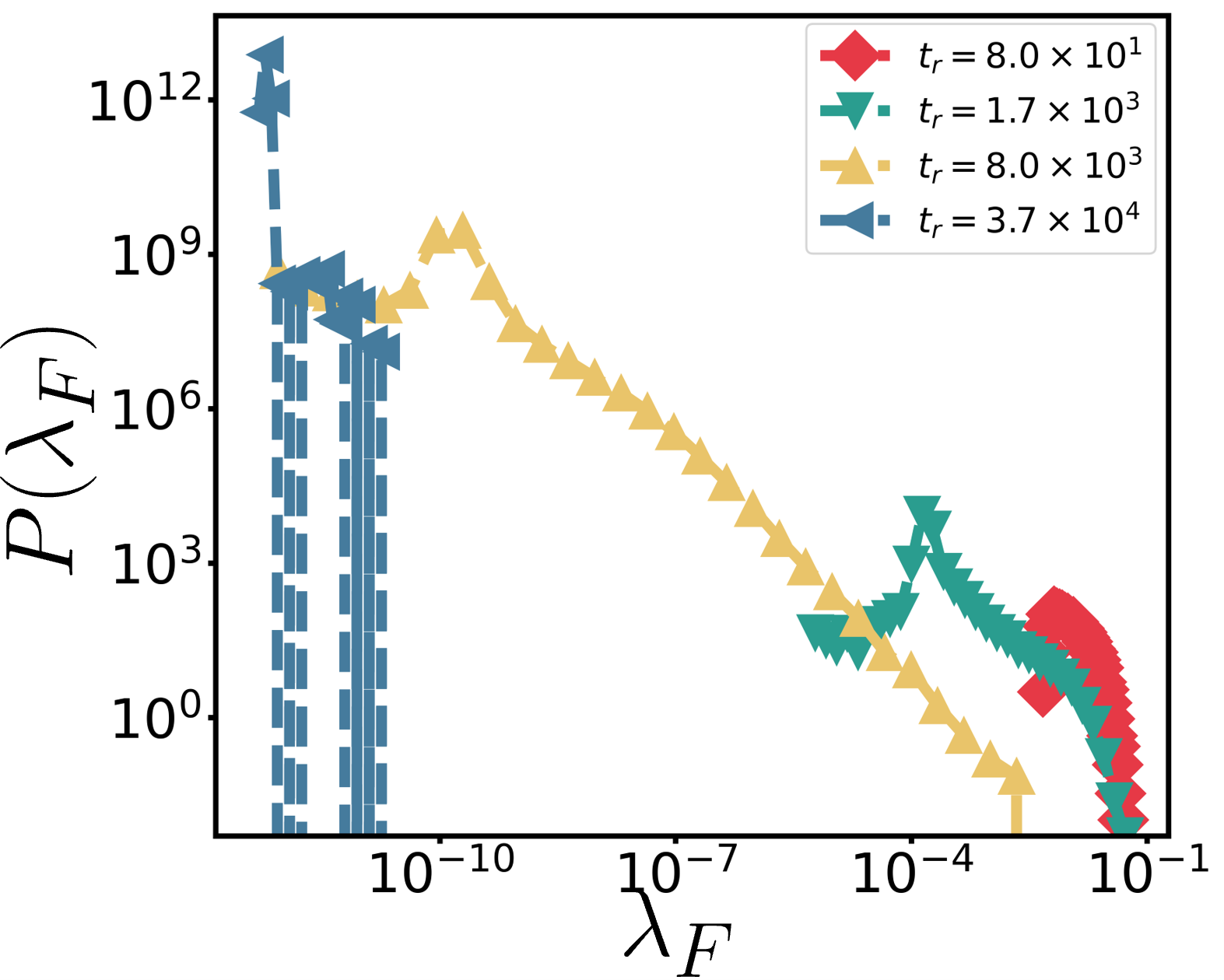}
\caption{\footnotesize \textbf{Approach to the fully elastic regime in $P(\lambda_F)$.} Probability distribution $P(\lambda_F)$ for increasing relaxation times $t_r$, using the Conjugate Gradient relaxation method under SS driving. As $t_r$ increases, the distribution first broadens over several decades, reflecting the coexistence of plastically active configurations and configurations that have ceased plastic activity. At the largest $t_r$, all configurations reach the elastic regime and the distribution becomes narrowly concentrated at the smallest resolvable residual forces, limited by machine precision. Simulations were performed with $N = 2048$ particles.  }
\label{fig:lambdaf_elastic_regime}
\end{figure}

Fig.~\ref{fig:lambdaf_elastic_regime} shows the evolution of the probability distribution $P(\lambda_F)$ for increasing relaxation times using the Conjugate Gradient relaxation method under SS driving. At short relaxation times, the distribution is centered at large values of $\lambda_F$, reflecting configurations that remain plastically active within the available relaxation time. Increasing $t_r$ leads to a broad distribution associated with the coexistence of configurations that have completed their plastic relaxation and configurations that remain active. For the largest relaxation times explored, the distribution becomes narrowly concentrated at low values of $\lambda_F$, near the numerical precision floor identified in the previous section. In this regime, plastic activity has ceased in essentially all configurations within the available relaxation time, and the residual forces approach the smallest values resolvable by the relaxation procedure.

\section{Interparticle Force Distributions Across Relaxation Times} 
\label{app:force_distrib}

To confirm that our SRF model does not introduce numerical artifacts associated with particle overlaps, we examine the distribution of interparticle force magnitudes $f_{ij}$ under SRF, using the Steepest Descent relaxation method and the same parameters as in the main text.

\begin{figure}[h]
\centering 
\includegraphics[width=0.45\textwidth]{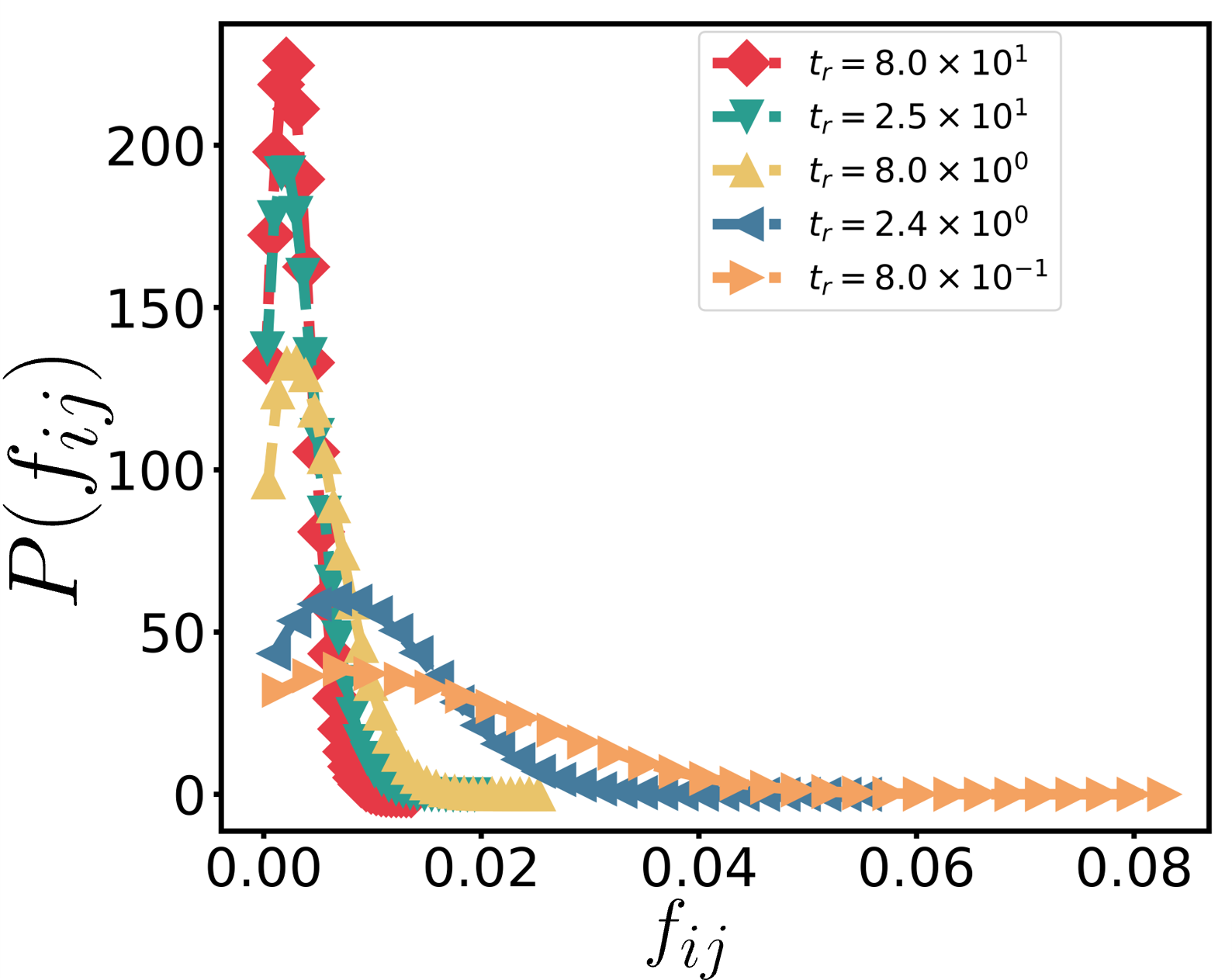}
\caption{\footnotesize \textbf{Distribution of interparticle force magnitudes under SRF driving for different relaxation times $t_r$.} As $t_r$ increases, the distributions shift towards lower force values and become progressively concentrated around a narrow peak, reflecting the increasing degree of relaxation achieved by the system. Simulations were performed using the Steepest Descent relaxation method with $N=8192$.  }
\label{fig:force_dist}
\end{figure}

Fig.~\ref{fig:force_dist} shows this distribution for several relaxation times $t_r$ spanning the range studied. The distribution exhibits a peak that shifts to lower $f_{ij}$ as $t_r$ increases, reflecting configurations progressively closer to mechanical equilibrium; unlike the residual force factor $\lambda_F$, this peak does not approach zero, since $f_{ij}$ measures the physical contact forces rather than the residual unbalanced force. Throughout the explored range, the distribution remains well below $f_{ij}\sim\epsilon/d_{ij}$, the force scale of order unity associated with complete particle overlap in our Hertzian definition, confirming that the sampled configurations contain no anomalously compressed contacts.

\newpage
\section{Results of conjugate gradients relaxation}
\label{app:CG_results}
While the main text focuses on results obtained with the Steepest Descent (SD) method, here we present the corresponding outcomes using Conjugate Gradient (CG) relaxation. Given the significantly lower computational cost of the CG method, results reported in this section were obtained from 30 independent realizations for each loading protocol.

\begin{figure}[h]
\centering 
\includegraphics[width=0.47\textwidth]{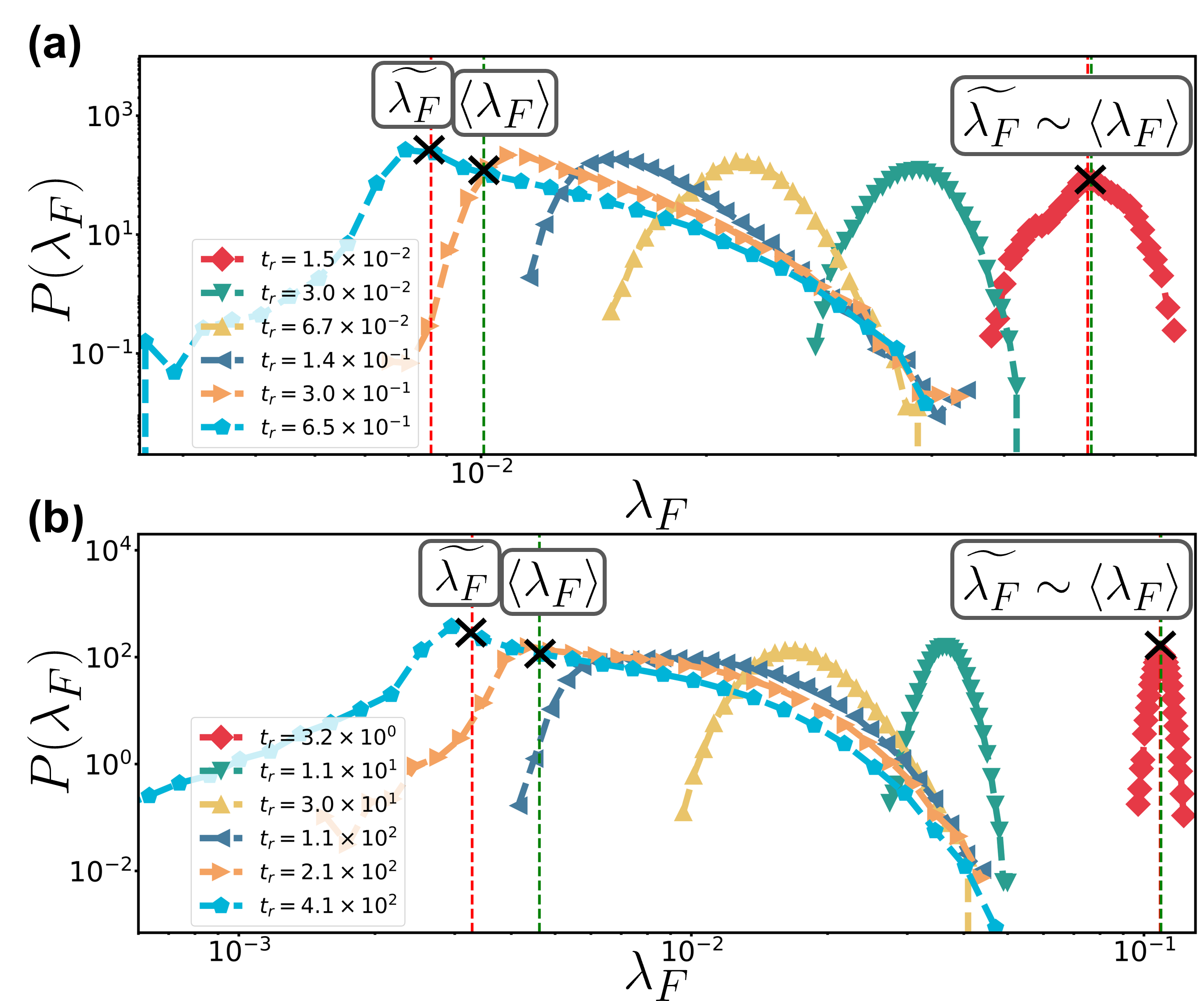}
\caption{\footnotesize \textbf{Probability distribution of $\lambda_F$ under Conjugate Gradient relaxation.}  $P(\lambda_F)$ for \textbf{(a)} SRF and \textbf{(b)} SS using the Conjugate Gradient relaxation method at different relaxation times $t_r$. Vertical lines represent, from left to right in each panel, the median $\tilde{\lambda_F}$ and the mean value $\langle \lambda_F \rangle$ for \textbf{(a)} $t_r = 3.1 \times 10^2$ and \textbf{(b)} $t_r = 1.9 \times 10^3$. The final pair of lines corresponds to the case where both values coincide, for \textbf{(a)} $t_r = 9.6$ and \textbf{(b)} $t_r = 63.0$. At low $t_r$, a single symmetric peak is observed, with $\tilde{\lambda_F} \! \approx\! \langle \lambda_F \rangle$. As $t_r$ increases, the distribution broadens and shifts, and at high $t_r$ a new dominant peak emerges at low $\lambda_F$ accompanied by a long tail. In this transitional regime, the median and mean differ ($\tilde{\lambda_F} \!\neq \!\langle \lambda_F \rangle$). Simulations were performed with $N = 8192$.}
\label{fig:CG_lambda_distributions} 
\end{figure}

\emph{Probability Distributions of $\lambda_F$.}-- The probability distributions obtained using Conjugate Gradient (CG) and Steepest Descent (SD) methods exhibit similar overall trends, with the transition being more pronounced in the CG case, as shown in Fig.~\ref{fig:CG_lambda_distributions}. At low $t_r$, the distribution displays a single, symmetric peak, with the median and mean nearly coinciding ($\tilde{\lambda_F} \!\approx\! \langle \lambda_F \rangle$): the relaxation time is too short for the system to reach an elastically relaxed state, so essentially all configurations remain plastically active. As $t_r$ increases, the distribution progressively broadens, reflecting the coexistence of two types of configurations: those in which avalanches are still relaxing plastically, at large $\lambda_F$, and those in which the system no longer reorganizes through avalanches and only relaxes its residual force, accumulating at low $\lambda_F$. This growing low-$\lambda_F$ population, together with the tail of still-active configurations, separates the median from the mean ($\tilde{\lambda_F} \!\neq\! \langle \lambda_F \rangle$).

\begin{figure}[h]
\centering 
\includegraphics[width=0.48\textwidth]{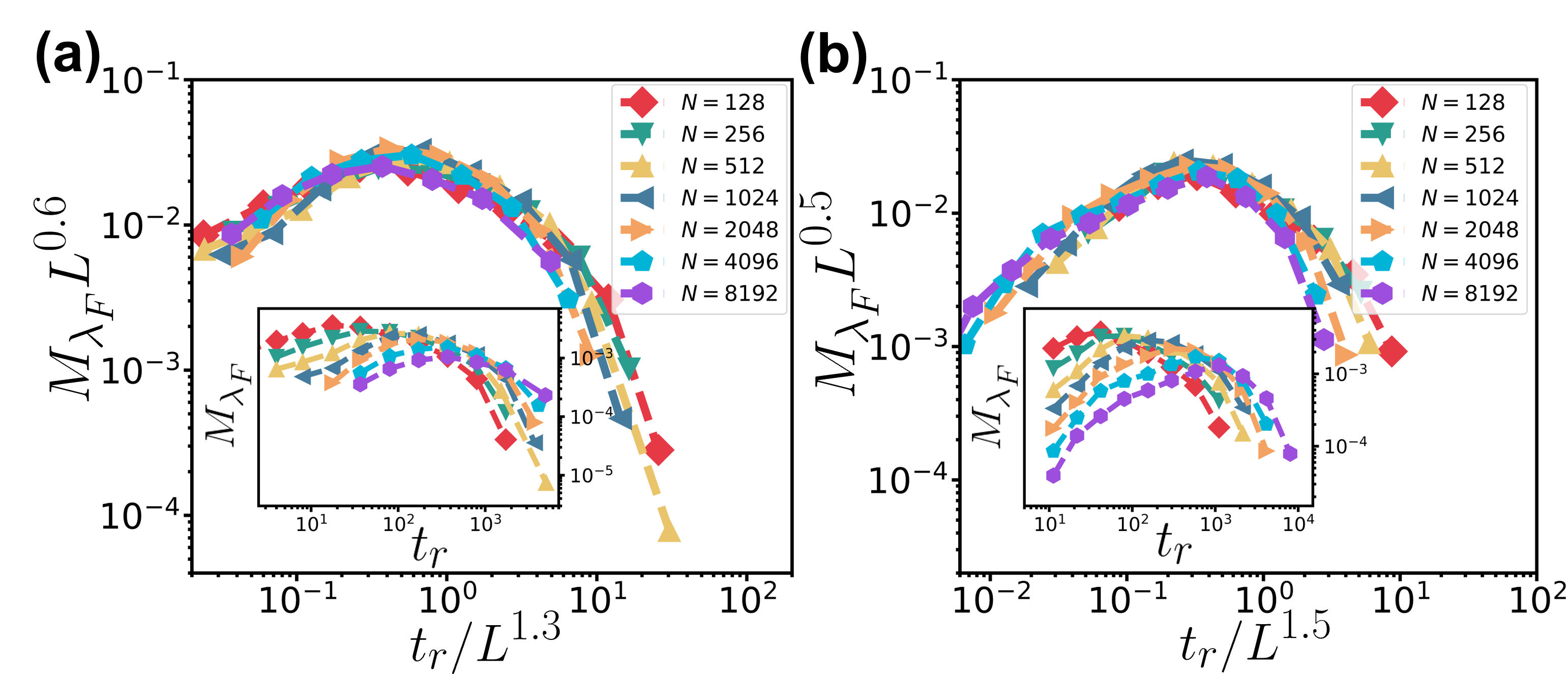}
\caption{\footnotesize \textbf{Collapse of $M_{\lambda_F}$ using Conjugate Gradient.} Finite-size collapsed $M_{\lambda_F}$ as a function of relaxation time $t_r$ for \textbf{(a)} SRF and \textbf{(b)} SS, using the Conjugate Gradient relaxation method. The insets display the raw data before rescaling. The extracted dynamic exponents are $z \!=\! 1.3$ for SRF and $z \!=\! 1.5$ for SS. All reported $z$ values correspond to the best fit, with values within $\pm 0.2$ remaining within a reasonable range.}
\label{fig:mlambda_CG}
\end{figure}

\emph{Dynamic exponent $z$.--} In Fig.~\ref{fig:mlambda_CG}, we present the finite-size collapse of $M_{\lambda_F}$ to extract the dynamic exponent $z$ for both SS and SRF under CG relaxation. a) presents $M_{\lambda_F}$ for SRF and b) for SS.

\begin{figure}[h]
\centering 
\includegraphics[width=0.49\textwidth]{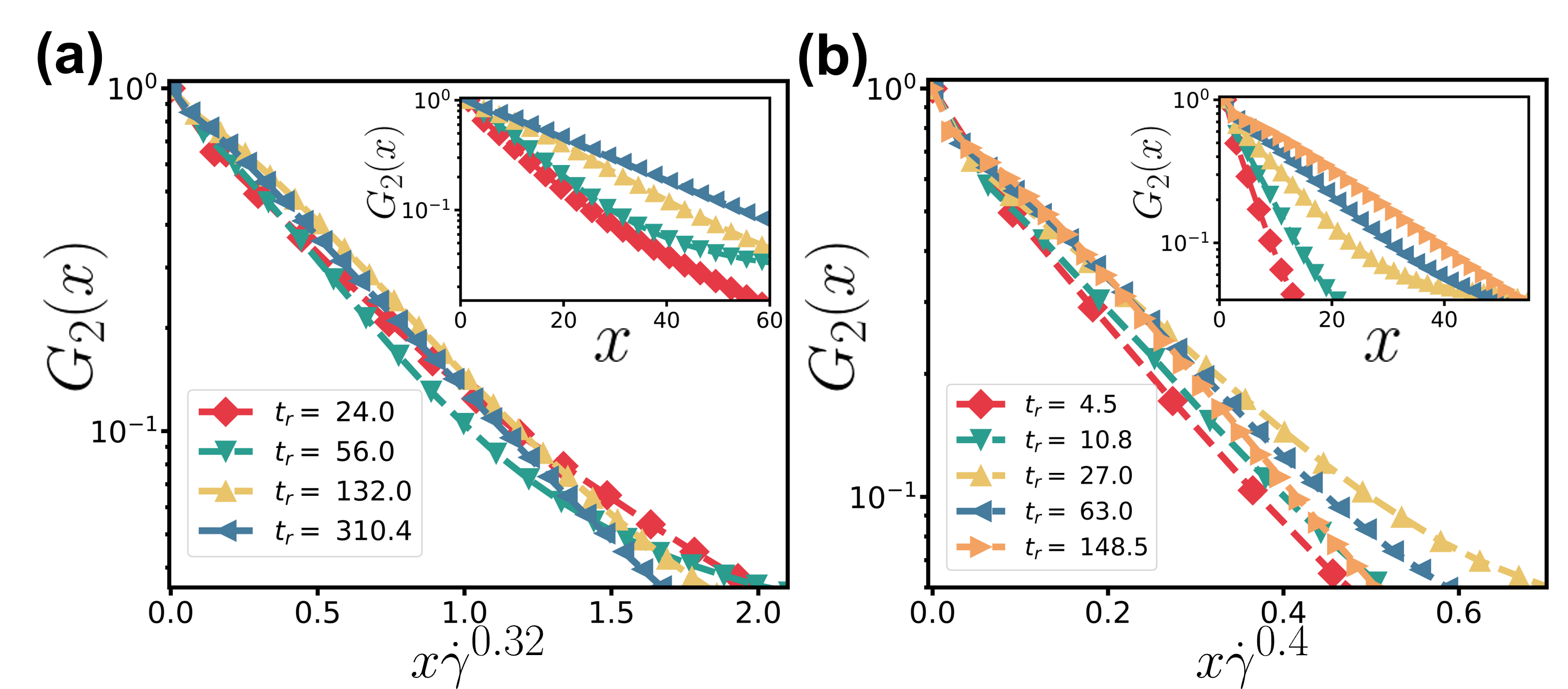}
\caption{\footnotesize \textbf{Scaling of $G_2(x)$ under Conjugate Gradient relaxation.}  Rescaled spatial correlation function $G_2(x)$ for \textbf{(a)} SRF and \textbf{(b)} SS, evaluated at various relaxation times $t_r$ using the Conjugate Gradient method. Insets display the unscaled data. We obtained $\nu / \beta = 0.32$ for SRF and $\nu / \beta = 0.4$ for SS. All reported $\nu / \beta$ values correspond to the best fit, with values within $\pm 0.03$ remaining within a reasonable range. Simulations were performed with $N = 8192$.}
\label{fig:correl_CG}
\end{figure}

\emph{Spatial correlation function}.-- We also compute and analyze the spatial correlation function $G_2(x)$ using CG, allowing a comparison with the results presented in the main text for SD. Fig.~\ref{fig:correl_CG} shows $G_2(x)$ for a) SRF and b) SS with the proper collapse.
\section{Independence from Step Size}
\label{app:step_size}
\begin{figure}[h]
\centering 
\includegraphics[width=0.44\textwidth]{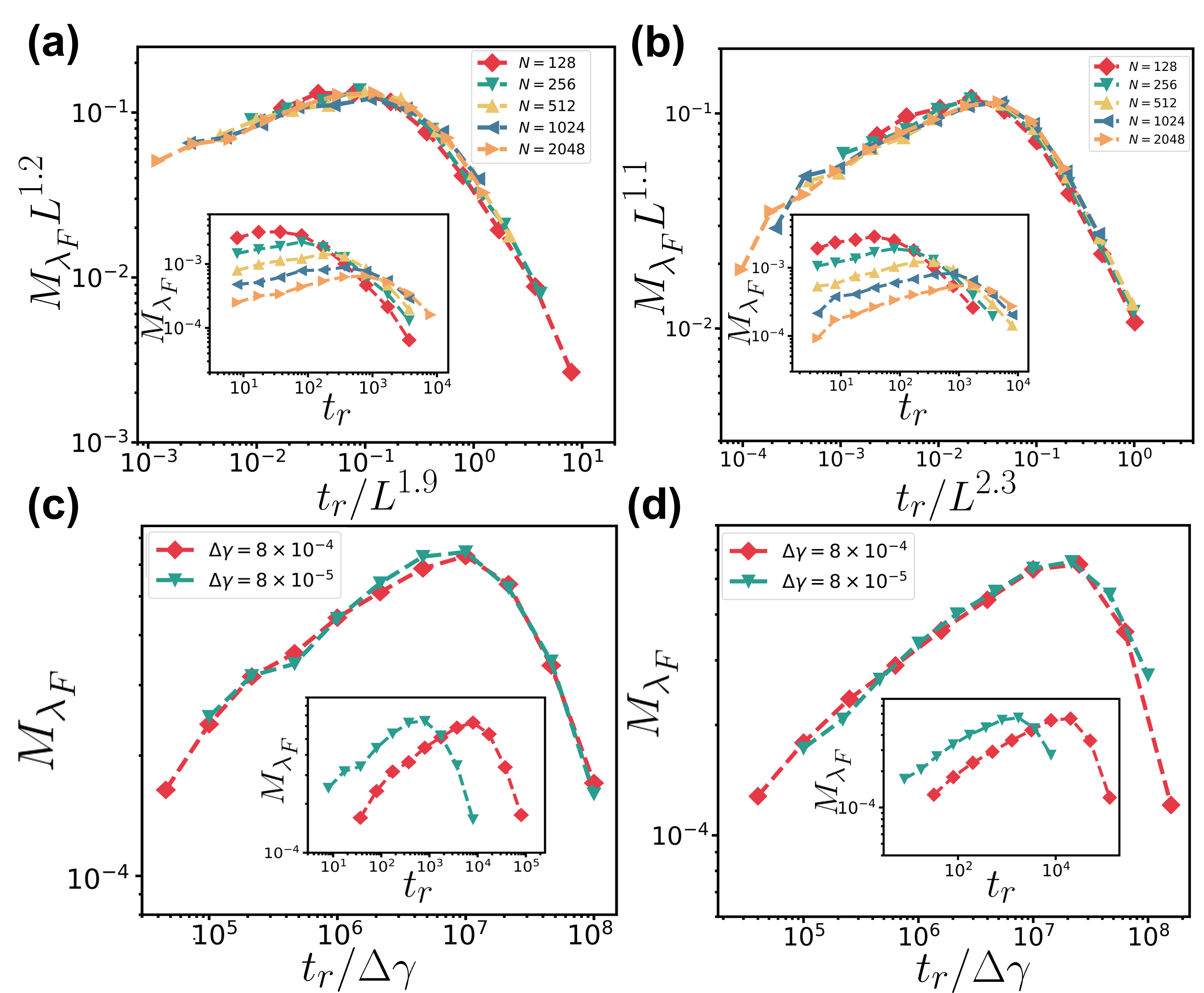}
\caption{\footnotesize \textbf{Finite-size scaling of $M_{\lambda_F}$ for different $\Delta \gamma$.} Panels \textbf{(a)} and \textbf{(b)} show finite-size collapsed $M_{\lambda_F}$ as a function of relaxation time $t_r$ for SRF and SS respectively, using the Steepest Descent relaxation method and $\Delta \gamma = 8 \times 10^{-5}$. Insets display the raw data before rescaling. The extracted dynamic exponents are $z \!=\! 1.9$ for SRF and $z \!=\! 2.3$ for SS, consistent with those obtained using the default strain increment $\Delta \gamma = 8 \times 10^{-4}$. Panels \textbf{(c)} and \textbf{(d)} show $M_{\lambda_F}$ vs.\ $t_r$ at fixed system size $N=4096$ for both $\Delta \gamma = 8 \times 10^{-4}$ and $\Delta \gamma = 8 \times 10^{-5}$, for SRF and SS respectively. The curves collapse upon rescaling $t_r$ by $\Delta \gamma$, confirming that the difference in relaxation time is solely determined by the strain increment.}
\label{fig:z_step_size_independence} 
\end{figure}

To verify that the extracted value of the dynamic exponent $z$ is not sensitive to the choice of $\Delta \gamma$, we computed $M_{\lambda_F}$ using an alternative strain increment, $\Delta \gamma = \Delta \gamma^{rnd} = 8 \times 10^{-5}$. As shown in Fig.~\ref{fig:z_step_size_independence}.a,b, the collapse remains consistent, and the extracted $z$ values fall within the previously reported uncertainty. This confirms that the determination of $z$ from the collapse of $M_{\lambda_F}$ is robust with respect to the specific choice of $\Delta \gamma$, as long as the system remains in the quasistatic regime.

To better understand the role of the strain increment $ \Delta \gamma $, we recall that after each simulation step, the applied stress perturbation scales with $ \Delta \gamma $ through the linear elastic response: $ \delta \sigma \sim G \Delta \gamma $, where $ G $ is the shear modulus. Consequently, the time required by a given relaxation method to reach a comparable state of mechanical equilibrium also scales with $ \Delta \gamma $. In this regime—where higher-order effects are negligible—the relation $ \dot{\gamma} = \Delta \gamma / t_r $ guarantees that different combinations of $ \Delta \gamma $ and $ t_r $ leading to the same $ \dot{\gamma} $ yield equivalent results.  As demonstrated in Fig.~\ref{fig:z_step_size_independence}.c,d, when comparing $ M_{\lambda_F} $ vs.\ $ t_r $ curves for two different values of $ \Delta \gamma $ at fixed system size, the curves collapse when rescaled by $ \Delta \gamma $. This indicates that the shift in the relaxation time at which avalanches reach the system size is solely determined by the difference in the applied strain increment.

\section{Correlation Function}
\label{app:correl_func}

The correlation function is defined as~\cite{CD1}:
\begin{equation*}
    G_2(x) = \langle | \delta \vec{\nu} (0) | | \delta \vec{\nu} (x) | \rangle - \langle | \delta \vec{\nu} (0) | \rangle \langle  | \delta \vec{\nu} (x) | \rangle,
    \label{eq:correlation}    
\end{equation*}

\noindent
where $\delta \vec{\nu}(x)$ denotes the non-affine velocity field. In the SS model, it is given by ${\delta \vec{\nu}_i \equiv \vec{\nu_i} - \dot{\gamma} (\vec{r_i} \cdot \hat{y})\hat{x}}$, while in the SRF model, it takes the form ${\delta \vec{\nu}_i \equiv \vec{\nu_i} - \vec{\nu}_{\parallel} {}^{rnd} \hat{n_i}^{rnd}}$. This correlation function captures how avalanches create localized velocity fluctuations, revealing spatial correlations in the deformation process. Since the focus is on velocity magnitudes rather than directions, the method effectively identifies regions of strong plastic activity regardless of the specific velocity orientation.

\section{Avalanche Distribution}
\label{app:aval_dis}
This section explores the significance of the avalanche statistics and the key exponents used in this work. The concepts are based on a previous study~\cite{D3SM01354E}. In this context, we use $\delta$ in the scaling relation presented in the main text, where it is defined as ${\delta = d - d_f (2 - \tau)}$.

This relation is based on the total stress released by an avalanche, given by $S = \delta \sigma L^d$, in the quasistatic limit. The size distribution follows a power law, ${P(S) \sim S^{-\tau}}$, with a cutoff $S_c$ imposed by the system size, given by ${S_c \sim L^{d_f}}$, where $d_f$ is the fractal dimension of the avalanche. Integrating $P(S)$ with the system-size-dependent cutoff yields ${\delta = d - d_f (2 - \tau)}$. Using the previously obtained values $d_f = 1.1$ for SS and $d_f = 1$ for SRF~\cite{D3SM01354E}, we can compute $\delta$ for both cases.


\section{Results Using a Divergent Interaction Potential} 
\label{app:lp_pot}

To further rule out numerical artifacts in the SRF model, we repeat the analysis using an interaction potential that diverges at full overlap. The Hertzian potential used in the main text remains finite as the interparticle distance vanishes, so overlaps are only softly penalized and could, in principle, influence the measured behavior. To test this, we performed simulations with the LP potential introduced by Lerner and Procaccia~\cite{PhysRevE.79.066109} (defined in Eq.~\ref{eq:lp_pot}, with $a_0=12$ and $b_0=6$), which diverges as the interparticle distance approaches zero and therefore prevents particle overlaps and crossings. This potential was employed in a previous work~\cite{D3SM01354E}. All simulation parameters and analysis procedures were kept identical to those in the main text, including the SRF deformation protocol and the Steepest Descent relaxation method. 

\begin{widetext} 
\begin{equation} 
U(r_{ij})= 
\begin{cases} \epsilon \left[ \left(\frac{d_{ij}}{r_{ij}}\right)^{a_0} -\frac{a_0(a_0+2)}{8} \left(\frac{b_0}{a_0}\right)^{\frac{a_0+4}{a_0+2}} \left(\frac{r_{ij}}{d_{ij}}\right)^4 +\frac{b_0(a_0+4)}{4} \left(\frac{r_{ij}}{d_{ij}}\right)^2 -\frac{(a_0+2)(a_0+4)}{8} \left(\frac{b_0}{a_0}\right)^{\frac{a_0}{a_0+2}} \right] & r_{ij}<d_{ij} \left(\frac{a_0}{b_0}\right)^{\frac{1}{a_0+2}} \\[10pt] 0 & r_{ij}>d_{ij} \left(\frac{a_0}{b_0}\right)^{\frac{1}{a_0+2}}  
\end{cases} 
\label{eq:lp_pot}
\end{equation} 
\end{widetext}

\begin{figure}[h]
\centering 
\includegraphics[width=0.48\textwidth]{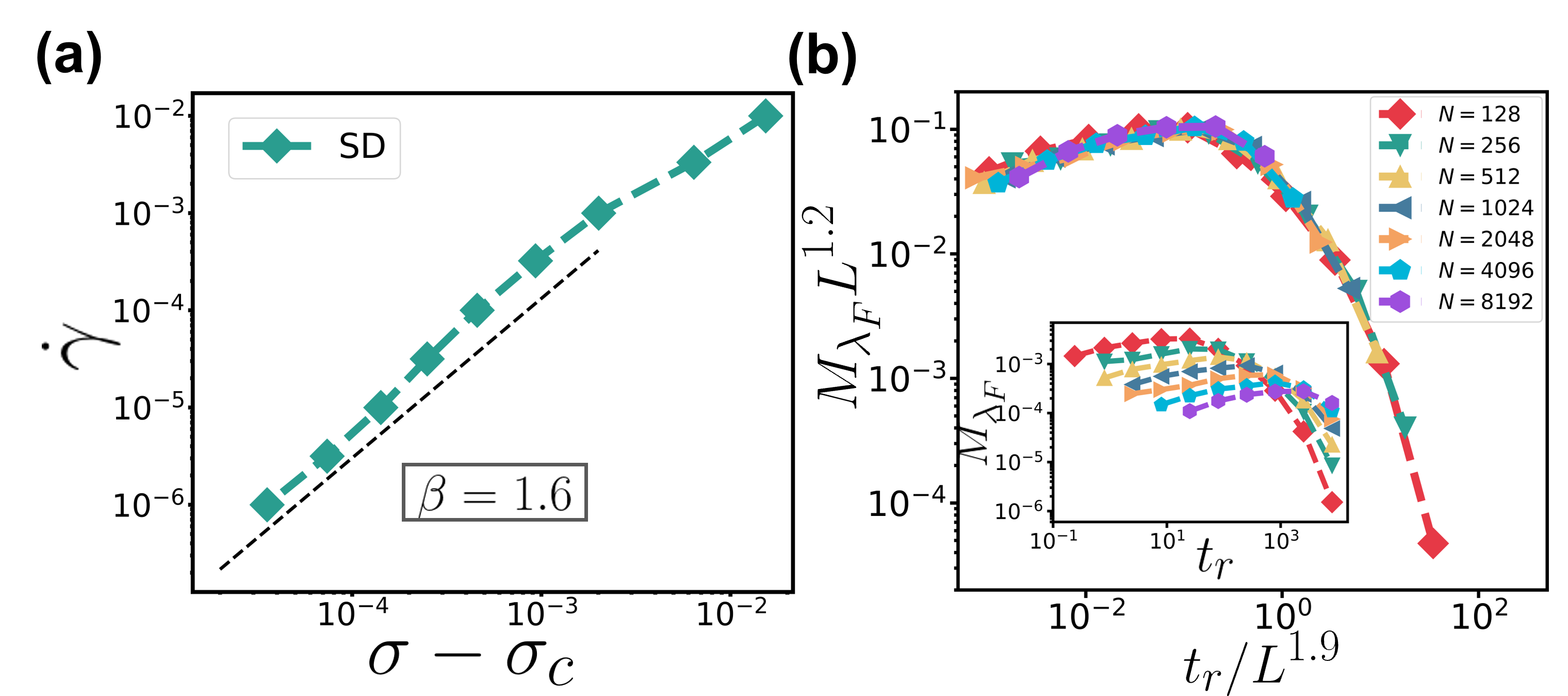}
\caption{\footnotesize \textbf{Results obtained using the LP interaction potential.} \textbf{(a)} Flow curves under SRF driving obtained with the Steepest Descent relaxation method for different system sizes. \textbf{(b)} Evolution of $M_{\lambda_F}$ as a function of the relaxation time $t_r$. The same rheological behavior and characteristic evolution of $M_{\lambda_F}$ reported for the Hertzian interaction are recovered with the LP potential. }
\label{fig:lp_pot}
\end{figure}

Fig.~\ref{fig:lp_pot}.a shows the flow curves obtained with the LP potential, which recover the rheological behavior reported in the main text throughout the explored range of strain rates. Fig.~\ref{fig:lp_pot}.b shows $M_{\lambda_F}$ as a function of the relaxation time; its evolution remains unchanged under the LP interaction, so that the dynamic exponent $z$ extracted from the finite-size collapse is preserved. Since a potential that excludes overlaps yields the same results, the SRF behavior is not affected by overlap-related artifacts.

\bibliography{referencias}